\documentclass[lettersize,journal]{IEEEtran}
\usepackage{amsmath,amsfonts}
\usepackage{mathtools}
\usepackage{empheq}
\usepackage{algorithmic}
\usepackage{array}
\usepackage[caption=false,font=normalsize,labelfont=sf,textfont=sf]{subfig}
\usepackage{textcomp}
\usepackage{stfloats}
\usepackage{url}
\usepackage{verbatim}
\usepackage{booktabs}
\usepackage{multirow}
\usepackage{graphicx}
\usepackage[normalem]{ulem}
\usepackage{xcolor}
\usepackage{color}
\usepackage{cite}
\usepackage[colorlinks=true, linkcolor=blue, urlcolor=black, citecolor=blue]{hyperref}
\usepackage{balance}
\usepackage{placeins}
\usepackage{float}

\def\BibTeX{{\rm B\kern-.05em{\sc i\kern-.025em b}\kern-.08em
    T\kern-.1667em\lower.7ex\hbox{E}\kern-.125emX}}

\begin{document}

\title{ARSAR-Net: Adaptively Regularized SAR Imaging Network with Efficient Unfolding}
\author{Shiping Fu, Yufan Chen, Zhe Zhang, Qixiang Ye
\thanks{Manuscript for arXiv submission. Shiping Fu, Yufan Chen, and Qixiang Ye are with the University of Chinese Academy of Sciences, Beijing, China. Zhe Zhang is with the Aerospace Information Research Institute of Chinese Academy of Sciences, Beijing, China.}}

\markboth{}%
{ARSAR-Net for SAR Imaging}

\maketitle
\begin{abstract}
Developed from sparse reconstruction approaches, deep unfolding networks (DUNs) have constituted an emerging method for synthetic aperture radar (SAR) imaging, offering fast convergence and data-driven learning.
However, baseline unfolding networks, derived from iterative sparse reconstruction algorithms such as alternating direction method of multipliers (ADMM), lack generalization capability across scenes, as their regularizers are empirically designed and keep unchanged during imaging.
In this study, we introduce a learnable regularizer to the unfolding network and propose an adaptively regularized SAR imaging network (ARSAR-Net) for scene-agnostic imaging (imaging across heterogeneous scenes of varying sparsity levels).
In practice, the vanilla ARSAR-Net suffers from inherent structural limitations in 2D signal processing, primarily due to its reliance on matrix inversion.
To conquer this, we further develop an ADMM without matrix inversion for efficient unfolding, by designing linear operations to replace the time-consuming matrix inversion operations. 
Experiments upon simulated and real-data demonstrate three advantages of ARSAR-Net: (1) a PSNR gain of up to 2.0 dB in imaging quality compared to existing deep network based imaging methods, (2) enhanced adaptability to complex scenes, and (3) a 50\% increase in imaging speed over existing unfolding networks.
These advancements establish a new paradigm for efficient and scene-agnostic SAR imaging systems.
Code is available at \href{https://github.com/ShipenFyu/ARSAR-Net}{\textit{github.com/ShipenFyu/ARSAR-Net}}.
\end{abstract}

\begin{IEEEkeywords}
Synthetic aperture radar imaging, Sparse microwave imaging, Adaptive regularization, Unfolding network, Deep learning
\end{IEEEkeywords}

\section{Introduction}
\label{sec:intro}

Synthetic aperture radar (SAR), an important microwave imaging technique, has been widely used in various earth observation applications such as flood observation, forestry research, and agriculture investigation~\cite{cumming2005digital}, thanks to its unique all-weather/all-time operational reliability and high-resolution imaging capacity~\cite{cumming2006improved, fu2022differentiable}. 
Different from other imaging approaches, an imaging process is needed to convert raw data into SAR image. Conventional widely-used SAR imaging techniques, such as the range-Doppler algorithm (RDA)~\cite{bamler1992comparison}, chirp scaling algorithm (CSA)~\cite{raney1994precision} and $\omega$-K algorithm \cite{cumming2003interpretations}, typically use matched filtering (MF) techniques. 
In recent decades, sparse signal processing, represented by compressed sensing (CS) theory, has been introduced into SAR imaging.
The derived sparse SAR imaging methods not only enjoy benefit of imaging from undersampled data, such as fewer measurements than those limited by the Nyquist theory, but also improve imaging performance, $e.g.$, lower sidelobes and better ambiguity suppression. To address the computational complexity issues of sparse reconstruction algorithms, approximated observation operators constructed from MF algorithms~\cite{fang2013fast} have also been adopted to achieve accelerated sparse SAR imaging. Despite the progress, generalization capability of sparse SAR imaging methods remains insufficient as hand-crafted regularizers lack adaptability to scenes with varying sparsity levels. 

\begin{figure*}[t]
    \centering
    \includegraphics[width=0.9\textwidth]{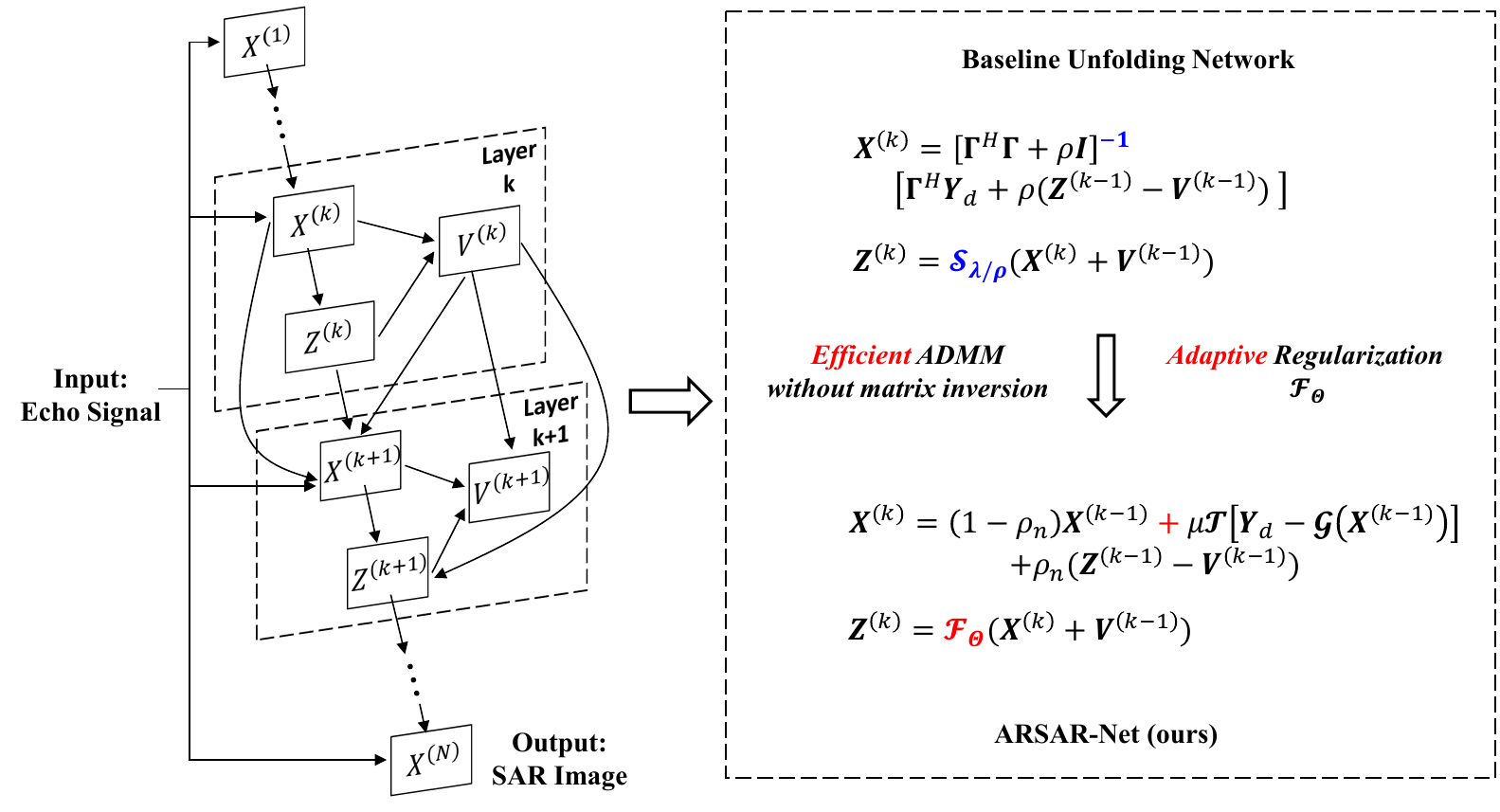}
    \caption{ARSAR-Net architecture. It overcomes the challenges of unfolding networks by introducing adaptive regularization and efficient unfolding.
    }
    \label{fig:shot}
\end{figure*}

Most recently, unfolding networks~\cite{monga2021algorithm} have emerged as a promising method for SAR imaging. As shown in Fig.~\ref{fig:shot}, these networks are constructed by iterative reconstruction algorithms, $e.g.$, the iterative shrinkage-thresholding algorithm (ISTA)~\cite{daubechies2004iterative} and the alternating direction method of multiplier (ADMM)~\cite{boyd2011distributed}, into a cascading neural network. 
In this way, the derived approaches take advantage of both model-driven algorithms and data-driven networks, achieving enhanced performance while enjoying interpretability. 

Nevertheless, the baseline unfolding networks remain challenged by scene generalization capability as the regularizers are also empirically designed.
Moreover, these networks are constrained by the structural limitations of the 2D signal processing, when relying on matrix inversion operation. 
Efficient alternatives to the time-consuming matrix inversion in the ADMM-based unfolding networks need to be further explored.

In this study, we propose a SAR imaging network with adaptive regularization, aimed at enhancing generalization across scenes of varying sparsity levels.
ARSAR-Net incorporates a pre-trained network as a learnable regularizer, replacing empirically designed priors and enabling scene-agnostic imaging.
The vanilla ARSAR-Net is built upon the ADMM algorithm for 2D signal processing, which suffers from inherent structural limitations.
To solve, we further introduce a linear approximation to replace the time-consuming matrix inversion in ADMM, thereby transforming the unfolding network into the more efficient ARSAR-Net.
In Fig.~\ref{fig:shot}, we show the framework of ARSAR-Net and its distinctions over the canonical unfolding network based SAR imaging approaches.

The contributions of this study are summarized as follows.
\begin{itemize}
    \item We develop an efficient ADMM algorithm for SAR imaging. The key is to replace the time-consuming matrix inversion of ADMM with an efficient linear operator derived from Taylor expansion. 

    \item Based on efficient ADMM, we design an efficient unfolding network (ARSAR-Net), which leverages a pre-trained deep network as the regularizer to improve the generalization capability of SAR imaging models across scenes.

    \item Experiments demonstrate the scene-agnostic capacity of ARSAR-Net. When handling real echo signals from various scenes, ARSAR-Net improves PSNR by up to 2.0 dB, while running 50\% faster than existing unfolding networks. 
\end{itemize}

\section{Related work}

\subsection{Sparse signal processing and sparse SAR imaging}

Compressed sensing-inspired regularization methods~\cite{baraniuk2007compressive, donoho2006compressed} have shown great potential when reconstructing sparse signals from limited observation data.
In these approaches, $\ell_1$-norm minimization~\cite{zhang2012sparse} was commonly formulated as a regularized inverse optimization problem, reconstructing target scenes by solving an objective function.
Considering the scattered field of SAR is sparse, iterative algorithms, such as orthogonal matching pursuit (OMP)~\cite{tropp2007signal}, iterative shrinkage-thresholding algorithm (ISTA), and fast iterative shrinkage-thresholding algorithm (FISTA)~\cite{beck2009fast} were applied.

While these improvements enable direct processing of downsampled raw data, they remain challenged by large-scale measurement matrices. To conquer this, approximated observation operators~\cite{fang2013fast} were proposed to reconstruct SAR images from downsampling data for high-quality imaging. 
An efficient $\ell_q$ regularizer~\cite{jiang2014efficient}, constructed using a decoupling operator based on the chirp scaling algorithm, reduced the computational complexity from quadratic to linear order.

Despite the effectiveness, these methods struggle to define a general sparse basis for varying sparsity levels. To address this, structural sparse representation~\cite{shen2014sar} formulated SAR image reconstruction and sparse space updating as a joint optimization problem. In addition, compound regularizer~\cite{xu2020accurate} was also proposed for adaptive sensing. 

While these methods address data storage and processing problems, they face two key challenges: (1) compressed sensing methods typically involve numerous hyperparameters that need to be manually tuned, and (2) their adaptability is limited by handcrafted priors and parameter dependencies.

\subsection{Deep neural network based SAR Imaging}
These have demonstrated remarkable effectiveness in signal and image processing, which inspires SAR imaging\cite{gregor2010learning}.
TPSSI-Net~\cite{wang2021tpssi}, a two-path deep unfolding network, was developed to integrate iteration-based reconstruction into a learnable framework, achieving fast and high-quality 3D SAR sparse imaging.
A deep SAR imaging network~\cite{pu2021deep} was proposed to jointly estimate scattering coefficients and compensate motion errors through an auto-encoder architecture, resulting in more robust and accurate imaging performance.

In contrast to DNNs, unfolding networks integrate iterative algorithms with deep learning~\cite{song2023dynamic}.
By mapping each iteration of the algorithm to a network layer, they improved imaging performance and interpretability.
For instance, AMP-Net~\cite{zhang2020amp} unfolded the AMP algorithm into a learnable deep architecture for compressive image sensing.
Building on this idea, NESTD-Net~\cite{gan2024nestd} incorporated a dual-path deblocking mechanism within a NESTA-inspired unfolding framework, further improving reconstruction fidelity and structural consistency.

Motivated by these studies, deep unfolding methods for SAR imaging have gained significant attention~\cite{wang2021tpssi}.
LRSR-ADMM-Net~\cite{an2022lrsr} was proposed to formulate SAR imaging as a joint low-rank and sparse recovery problem within an unfolded ADMM framework.
For non-sparse scenes, SR-CSA-Net~\cite{zhang2023nonsparse}, designed as ISTA-based unfolding networks that employ CSA-generated imaging operators.
Additionally, NSR-Net~\cite{yang2024sar} employed a neural network to learn a representation by approximating the iterative solution obtained via proximal gradient descent (PGD). 
Nevertheless, despite improved learning capabilities, computational efficiency remains a significant challenge.

\subsection{Learnable regularization}

Learnable regularization can be implemented via denoisers or neural networks, serving as a data-driven alternative to conventional regularizers.
A representative approach is the plug-and-play prior ~\cite{venkatakrishnan2013plug}, which replaced the conventional regularizer with a denoiser for magnetic resonance imaging (MRI) reconstruction. 
ADMM-CSNet~\cite{yang2018admm} embedded a CNN-based optimization module with nonlinear piecewise functions into ADMM, achieving superior performance and efficiency in MRI/optical image reconstruction.
Inspired by these advances, ARSAR-Net is developed in this study to further explore the potential of a learnable regularizer for SAR imaging.

\section{Efficient ADMM for SAR imaging}
\label{sec:admm}

For self-containing, we first introduce a 2D SAR imaging model based on CSA operators upon ADMM. 
We then propose an efficient ADMM without matrix inversion for 2D echo signal processing.

\subsection{2D SAR Imaging with ADMM}

\begin{figure}[t]
    \centering
    \includegraphics[width=0.7\columnwidth]{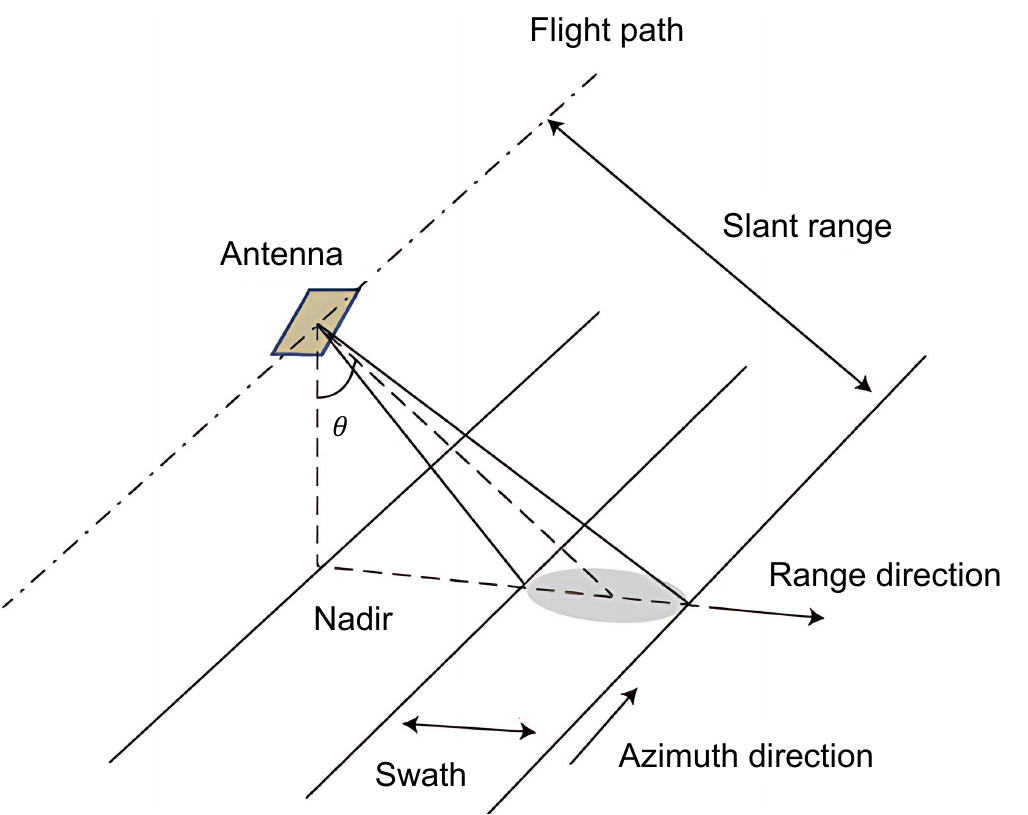}
    \caption{Geometric Structure of SAR Imaging System(Credit: NASA).~\cite{RadarGeometry}}
    \label{radar}
\end{figure}

The vertical side-looking strip mode is employed in the SAR imaging, Fig.~\ref{radar}. 
The SAR imaging model establishes a mathematical relationship between the radar echo signal and the scattering coefficients of the observed area.
For numerical processing, the observation area is discretized to a grid of $M \times N$ scattering centers, where the indices $m = 1, 2, \dots, M$ and $n = 1, 2, \dots, N$ denote discrete positions along the range and azimuth directions, respectively.
This discretization allows the scattering coefficients to be denoted as a 2D matrix $\boldsymbol{X} \in \mathbb{C}^{M \times N}$. Similarly, the echo signal is represented as $\boldsymbol{Y} \in \mathbb{C}^{M \times N}$. 

To characterize the relationship between them, the echo signal is modeled by the observation matrix $\mathbf{\Phi} \in \mathbb{C}^{MN \times MN}$ which acts on the vectorized scattering coefficient $\boldsymbol{x}=\mathrm{vec}(\boldsymbol{X})$ to produce the vectorized echo signal $\boldsymbol{y}=\mathrm{vec}(\boldsymbol{Y})$, $i.e.$, 
\begin{equation}
    \boldsymbol{y} = \mathbf{\Phi}\boldsymbol{x}
    \label{eq:vec}
\end{equation}
where $\mathrm{vec}(\cdot)$ denotes the vectorization operator stacking the matrix columns into a single vector. 
To simulate reduced observations, the echo signal is further downsampled by applying $\boldsymbol{\Psi}_r \in \mathbb{R}^{M' \times M}$ (range) and $\boldsymbol{\Psi}_a \in \mathbb{R}^{N' \times N}$ (azimuth), yielding the downsampled echo $\boldsymbol{Y_d} \in \mathbb{C}^{M' \times N'}$.

In Eq.~\ref{eq:vec}, vectorization leads to a very large observation matrix.
Converting an $N \times N$ echo signal into a vector results in a $N^2 \times N^2$ observation matrix $\mathbf{\Phi}$, substantially increasing storage requirements and computational costs.
Moreover, the inherent azimuth-range coupling further complicates the direct decoupling of the observation matrix.

To address these issues, the matched filter method~\cite{fang2013fast} approximates $\mathbf{\Phi}$ using a series of 1D operators along the azimuth and range directions. These operators are efficiently implemented via the chirp scaling algorithm (CSA)~\cite{dong2013novel}, which directly constructs the imaging operator $\mathcal{M}(\cdot)$ and its approximate inverse, the observation operator $\mathcal{H}(\cdot)$, as
\begin{equation}
\begin{split}
    \mathcal{M}(\boldsymbol{Y}) &= \left\langle\boldsymbol{F_r^H}\left\{\boldsymbol{F_r}\left[\left(\boldsymbol{Y}\boldsymbol{F_a}\right)\circ\mathbf{\Theta_s}\right]\circ\mathbf{\Theta_r}\right\}\circ\mathbf{\Theta_a}\right\rangle\boldsymbol{F_a^H}, \\
    \mathcal{H}(\boldsymbol{X}) &= \left\langle\boldsymbol{F_r^H}\left\{\boldsymbol{F_r}\left[\left(\boldsymbol{X}\boldsymbol{F_a}\right)\circ\mathbf{\Theta_a^*}\right]\circ\mathbf{\Theta_r^*}\right\}\circ\mathbf{\Theta_s^*}\right\rangle\boldsymbol{F_a^H},
\end{split}
\end{equation}
where $\circ$ denotes the Hadamard product. $\boldsymbol{F_r}$ and $\boldsymbol{F_a}$ are the FFT matrices along the range and azimuth directions, respectively. $\boldsymbol{F_r^H}$ and $\boldsymbol{F_a^H}$ denote the IFFT matrices. $\mathbf{\Theta_s}$, $\mathbf{\Theta_r}$, and $\mathbf{\Theta_a}$ are the quadratic phase function for chirp scaling operation, the phase function for range compression and bulk range cell migration correction (RCMC), and the phase function for azimuth compression and residual phase compensation, respectively.

By leveraging the CSA-based operators, the large observation matrix $\mathbf{\Phi}$ is effectively replaced, enabling the reconstruction of the scattering coefficient $\boldsymbol{X}$ from the measured echo signal $\boldsymbol{Y}$. 
To solve this inverse problem, we adopt the alternating direction method of multipliers (ADMM), a robust optimization algorithm widely used in regularized reconstruction.
By introducing dual variables, ADMM decomposes the problem into separable subproblems, allowing flexible handling of various regularization terms. 
Furthermore, ADMM incorporates a quadratic penalty term in the augmented Lagrangian, which enhances convergence~\cite{tomioka2009dual}.

Through ADMM, the reconstruction problem is decomposed into two separable parts involving $\boldsymbol{X}$ and auxiliary variable $\boldsymbol{Z}$, as
\begin{equation}
    \begin{array}{c}
    \underset{\boldsymbol{X}}{\min} \left\|\mathcal{G}(\boldsymbol{X}) - \boldsymbol{Y_d}\right\|_2^2 + \lambda \mathbf{\phi}(\boldsymbol{Z}) \\
    \text{s.t. } \boldsymbol{Z} = \boldsymbol{X},
    \end{array}
\end{equation}
where $\mathcal{G}(\cdot) = \mathbf{\Psi}_r\mathcal{H}(\cdot)\mathbf{\Psi}_a$ denotes the observation operator incorporating the downsampling matrices $\mathbf{\Psi}_r$ and $\mathbf{\Psi}_a$, $\lambda$ is the regularization factor, and $\mathbf{\phi}(\cdot)$ represents an arbitrary regularization function such as $\ell_1$ norm or TV regularization.

Accordingly, an augmented Lagrangian function is constructed, where the variable $\boldsymbol{U}$ is the dual variable and $\rho$ is the penalty parameter, as
\begin{equation}
\begin{split}
    L_{\rho}(\boldsymbol{X},\boldsymbol{Z},\boldsymbol{U}) =& \left\|\boldsymbol{Y_d} - \mathcal{G}(\boldsymbol{X})\right\|_2^2 + \lambda \mathbf{\phi}(\boldsymbol{Z}) \\
    &+ \frac{\rho}{2}\|\boldsymbol{Z} - \boldsymbol{X}\|_2^2 + <\boldsymbol{U}, \boldsymbol{Z} - \boldsymbol{X}>.
\end{split}
\label{eq:2D_Lagrangian}
\end{equation}

From Eq.~\ref{eq:2D_Lagrangian}, the $k$-th iteration of ADMM is derived by decomposing the problem into three sequential updates, as 
\begin{subequations}
\begin{empheq}[left=\empheqlbrace]{align}
\boldsymbol{X}^{(k)} =& \,\underset{\boldsymbol{X}}{\arg\min} \left\|\boldsymbol{Y_d} - \mathcal{G}(\boldsymbol{X})\right\|_2^2 \nonumber\\
&+ \frac{\rho}{2}\|\boldsymbol{X} - \boldsymbol{Z}^{(k-1)} + \boldsymbol{V}^{(k-1)}\|_2^2 \label{eq:2d x}\\[1ex]
\boldsymbol{Z}^{(k)} =& \,\underset{\boldsymbol{Z}}{\arg\min}\,\lambda \mathbf{\phi}(\boldsymbol{Z}) \nonumber \\
&+ \frac{\rho}{2}\|\boldsymbol{X}^{(k)} - \boldsymbol{Z} + \boldsymbol{V}^{(k-1)}\|_2^2 \label{eq:2d z}\\[1ex]
\boldsymbol{V}^{(k)} =& \,\boldsymbol{V}^{(k-1)} + \boldsymbol{X}^{(k)} - \boldsymbol{Z}^{(k)} \label{eq:2d v}
\end{empheq}
\end{subequations}
where $\boldsymbol{V} = 1/\rho\cdot\boldsymbol{U}$ is the scaled dual variable.

\begin{figure*}[!b]
    \hrule
    \vspace{0.5em}

    \begin{equation}
    \mathcal{G}(\boldsymbol{X}) = \mathbf{\Psi_r}\left\langle\boldsymbol{F_r^H}\left\{\boldsymbol{F_r}\left[\left(\boldsymbol{X}\boldsymbol{F_a}\right)\circ\mathbf{\Theta_a^*}\right]\circ\mathbf{\Theta_r^*}\right\}\circ\mathbf{\Theta_s^*}\right\rangle\boldsymbol{F_a^H}\mathbf{\Psi_a} 
    \label{eq:inverse imaging operator}
    \end{equation}
    
    \begin{equation}
    \mathcal{T}(\boldsymbol{Y}) = \left\langle\boldsymbol{F_r^H}\left\{\boldsymbol{F_r}\left[\left(\mathbf{\Psi_r}^H\boldsymbol{Y}\mathbf{\Psi_a}^H\boldsymbol{F_a}\right)\circ\mathbf{\Theta_s}\right]\circ\mathbf{\Theta_r}\right\}\circ\mathbf{\Theta_a}\right\rangle\boldsymbol{F_a^H}
    \label{eq:imaging operator}
    \end{equation}
\end{figure*}

Suppose that $\mathcal{G}(\cdot)$ in Subproblem~\ref{eq:2d x} is represented as a single matrix $\mathbf{\Gamma}$, the subproblem admits an explicit iterative solution, as
\begin{equation}
\boldsymbol{X}^{(k)} = \left[\mathbf{\Gamma}^H\mathbf{\Gamma} + \rho \mathbf{I}\right]^{-1} \left[\mathbf{\Gamma}^H \boldsymbol{Y_d} + \rho(\boldsymbol{Z}^{(k-1)} - \boldsymbol{V}^{(k-1)})\right],
\label{eq:normal solution of X}
\end{equation}
where $\mathbf{I}$ denotes the identity matrix.

However, the observation operator $\mathcal{G}(\cdot)$, formulated in Eq.~\ref{eq:inverse imaging operator}, combines matrix multiplications and Hadamard products, resulting in a complex mathematical structure.
Therefore, the 2D imaging model cannot derive an explicit iterative solution of scattering coefficient $\boldsymbol{X}$ via matrix inversion.
As a result, the conventional ADMM is ill-suited for 2D SAR imaging.

\subsection{ADMM without Matrix Inversion}
\label{sec:matrix-free_imaging}
This is proposed to resolve the incompatibility between conventional ADMM and 2D SAR imaging.
%
The solution methods for Subproblem $\boldsymbol{X}$ in Eq.~\ref{eq:2d x} and Subproblem $\boldsymbol{Z}$ in Eq.~\ref{eq:2d z} are derived in what follows.

\subsubsection{Subproblem X}

We employ a strategy that replaces matrix inversion with a Taylor expansion-based linear approximation to solve the subproblem $\boldsymbol{X}$ of 2D echo signal. Several alternatives to the conventional ADMM~\cite{wu2023learning, wang20223} have been derived from different perspectives and similarly arrive at solutions without matrix inversion.

Let $\boldsymbol{S}(\boldsymbol{X})$ denote the objective function to be optimized in the subproblem $\boldsymbol{X}$ in Eq.~\ref{eq:2d x}, which is formulated as
\begin{equation}
\begin{split}
    \boldsymbol{S}(\boldsymbol{X}^{(k)}) =& \,\left\lVert \mathcal{G}(\boldsymbol{X}^{(k)})-\boldsymbol{Y_d}\right\rVert_2^2 \\ &+ \frac{\rho}{2}\left\lVert \boldsymbol{X}^{(k)} - \boldsymbol{Z}^{(k-1)}+\boldsymbol{V}^{(k-1)}\right\rVert_2^2.
\end{split}
\end{equation}

We carry out a second-order Taylor expansion of the objective function $\boldsymbol{S}(\boldsymbol{X})$ while truncating higher-order terms.
\[
\begin{split}
\boldsymbol{X}^{(k)} =& \,\underset{\boldsymbol{X}}{\arg\min}\,\boldsymbol{S}(\boldsymbol{X}) \\
\approx& \,\underset{\boldsymbol{X}}{\arg\min}\boldsymbol{S}(\boldsymbol{X}^{(k-1)})+\nabla\boldsymbol{S}(\boldsymbol{X}^{(k-1)})^H(\boldsymbol{X} - \boldsymbol{X}^{(k-1)}) \\
&+\frac{1}{2}(\boldsymbol{X} - \boldsymbol{X}^{(k-1)})^H\nabla ^2\boldsymbol{S}(\boldsymbol{X}^{(k-1)})(\boldsymbol{X} - \boldsymbol{X}^{(k-1)}),
\end{split}
\]
where the second-order gradient matrix $\nabla ^2\boldsymbol{S}(\boldsymbol{X}^{(k-1)})$ refers to the Hessian matrix~\cite{van1994complex}.

Since the observation operator $\mathcal{G}(\cdot)$ in $\boldsymbol{S}(\boldsymbol{X})$ is approximately linear under practical radar system configurations, the Hessian matrix can be reasonably approximated as constant, which is thoroughly explained in Appendix~\ref{app:lipschitz analysis}, and expressed as
\begin{equation}
    \begin{split}
    \boldsymbol{X}^{(k)} \approx& \,\underset{\boldsymbol{X}}{\arg\min}\boldsymbol{S}(\boldsymbol{X}^{(k-1)})+\nabla\boldsymbol{S}(\boldsymbol{X}^{(k-1)})^H(\boldsymbol{X} - \boldsymbol{X}^{(k-1)}) \\
    &+\frac{L}{2}\left\lVert \boldsymbol{X} - \boldsymbol{X}^{(k-1)}\right\rVert_2^2,
    \end{split}
\label{eq:admm taylor}
\end{equation}
where $L$ denotes the Lipschitz constant, thus simplifying the iteration.

Through derivation of the optimization problem Eq.~\ref{eq:admm taylor}, we have the closed-form solution~\footnote{Please refer to Appendix~\ref{app:admm proof} for the detailed derivation.}, as 
\begin{equation}
\begin{aligned}
\boldsymbol{X}^{(k)} =& (1 - \rho_n)\boldsymbol{X}^{(k-1)} + \mu\mathcal{T}\left[\boldsymbol{Y_d} - \mathcal{G}(\boldsymbol{X}^{(k-1)})\right] \\
& +\rho_n(\boldsymbol{Z}^{(k-1)}-\boldsymbol{V}^{(k-1)}),  
\end{aligned}
\label{eq:admm solution}
\end{equation}
where $\rho_n = \rho/L$ is the normalized penalty parameter, $\mu = 2/L$ is a step size parameter derived from the Lipschitz constant, and $\mathcal{T}(\cdot)$ denotes the imaging operator with downsampling matrices, which can be formulated as Eq.~\ref{eq:imaging operator}.

\subsubsection{Subproblem Z}
\begin{figure}[t]
    \centering
    \includegraphics[width=0.7\columnwidth]{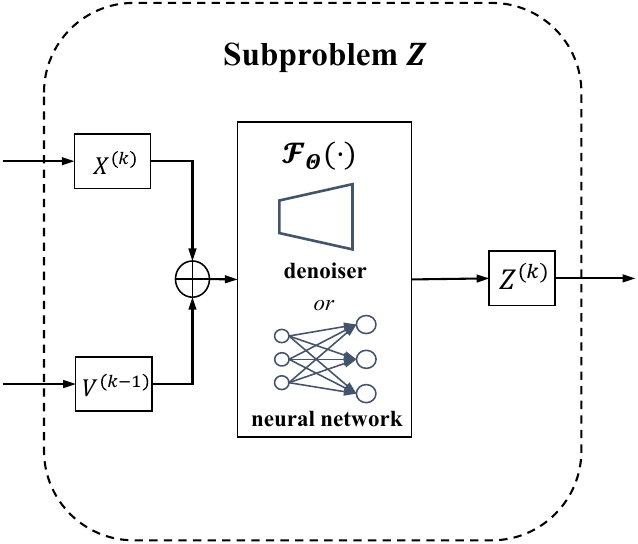}
    \caption{Computational diagram of subproblem Z}
    \label{fig:z_module}
\end{figure}

Replacing the empirical regularizer ($e.g.$, $\ell_1$-norm) with a function $\mathbf{\phi}(\cdot)$, the nonlinear proximal operator $\mathcal{F}(\cdot)$ determined by $\mathbf{\phi}(\cdot)$ is implemented through either model-based denoiser or data-driven neural networks.
Consequently, the closed-form solution of subproblem $\boldsymbol{Z}$ is formulated as
\begin{equation}
    \boldsymbol{Z}^{(k)} = \mathcal{F}_{\Theta}(\boldsymbol{X}^{(k)} + \boldsymbol{V}^{(k-1)}),
\end{equation}
where $\mathcal{F}_{\Theta}(\cdot)$ is the nonlinear proximal operator, and ${\Theta}$ denotes the parameters of the denoiser or neural network.
The computational flowchart of Subproblem $\boldsymbol{Z}$ is illustrated in Fig.~\ref{fig:z_module}.

\subsubsection{Efficient SAR Imaging}
By integrating the solutions of these subproblems, we obtain an efficient ADMM for SAR imaging, as 

\begin{subequations}
\small
\begin{empheq}[left=\empheqlbrace]{align}
\boldsymbol{X}: \boldsymbol{X}^{(k)} =& \,(1 - \rho_n)\boldsymbol{X}^{(k-1)}+ \mu\mathcal{T}\left[\boldsymbol{Y_d} - \mathcal{G}(\boldsymbol{X}^{(k-1)})\right] \nonumber \\
&+\rho_n(\boldsymbol{Z}^{(k-1)}-\boldsymbol{V}^{(k-1)}) \label{eq:nm x}\\[1ex]
\boldsymbol{Z}: \boldsymbol{Z}^{(k)} =& \,\mathcal{F}_{\Theta}(\boldsymbol{X}^{(k)} + \boldsymbol{V}^{(k-1)}) \label{eq:nm z}\\[1ex]
\boldsymbol{V}: \boldsymbol{V}^{(k)} =& \,\boldsymbol{V}^{(k-1)} + \boldsymbol{X}^{(k)} - \boldsymbol{Z}^{(k)} \label{eq:nm v}
\end{empheq}
\end{subequations}

The SAR imaging model, built upon efficient ADMM, overcomes the computational bottlenecks of conventional imaging frameworks, which lays a foundation for the ARSAR-Net.

\section{ARSAR-Net: Adaptively Regularized SAR Imaging Network}

\subsection{Efficient Unfolding Architecture}

\label{sec:net_arc}

\begin{figure*}[t]
    \centering
    \includegraphics[width=0.95\textwidth]{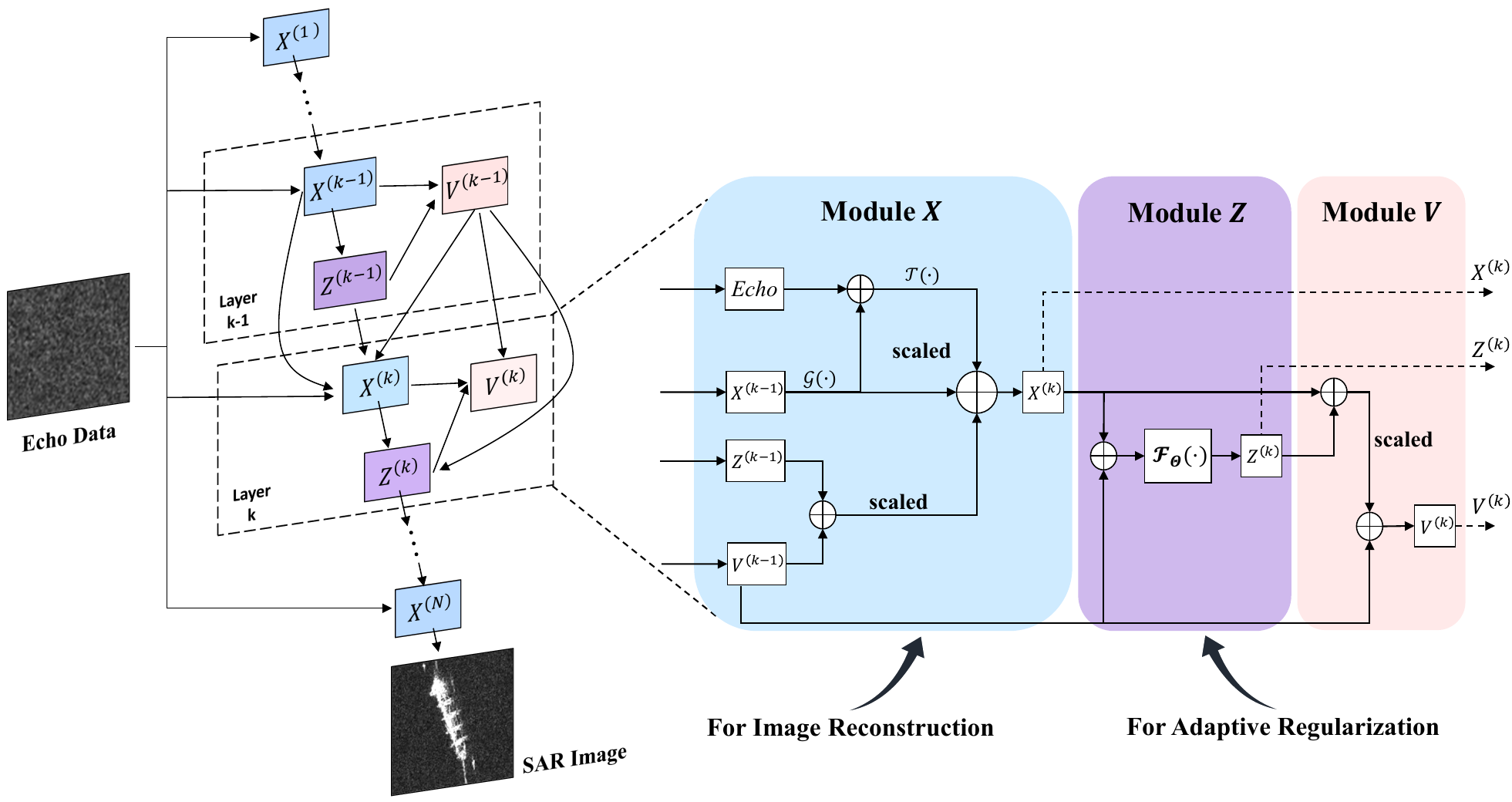}
    \caption{Architecture of ARSAR-Net, which consists of a reconstruction module ($\boldsymbol{X}$), an adaptive regularizer ($\boldsymbol{Z}$), and a scaled multiplier ($\boldsymbol{V}$). With the proposed efficient ADMM, ARSAR-Net is an efficient unfolding network with adaptive regularization.}
    \label{fig:overall}
\end{figure*}

Deep unfolding networks bridge model-based optimization and data-driven learning by mapping iterative algorithms to neural networks, enabling interpretable and efficient inference. 
Based on the efficient ADMM in Sec.~\ref{sec:matrix-free_imaging}, we design an efficient unfolding network, termed ARSAR-Net, where each layer corresponds to an iteration step of the ADMM. 
The designed unfolding network enables data-driven parameter optimization based on SAR echo data and corresponding ground-truth images.
As shown in Fig.~\ref{fig:overall}, the ARSAR-Net takes 2D echo data $\boldsymbol{Y_d}$ as input and outputs the scattering coefficient matrix $\boldsymbol{X}^{(N)}$, whose magnitude represents the SAR image. 
The network comprises $N_s$ layers, each containing three differentiable modules aligned with the variables in the 2D imaging model, detailed as follows.

\subsubsection{Reconstruction Module}
This module reconstructs the scattering coefficient matrix $\boldsymbol{X}$ by incorporating CSA operators that encode SAR imaging physics.
The computational process of the module is
\begin{equation}
\begin{split}
\boldsymbol{X}^{(k)} =& \,(1 - \widetilde{\rho})\boldsymbol{X}^{(k-1)}+ \widetilde{\mu}\mathcal{T}\left[\boldsymbol{Y_d} - \mathcal{G}(\boldsymbol{X}^{(k-1)})\right] \\
&+ \widetilde{\rho}(\boldsymbol{Z}^{(k-1)}-\boldsymbol{V}^{(k-1)}),
\end{split}
\label{eq:arsar-net module1}
\end{equation}
where:
\begin{itemize}
\item $\widetilde{\rho}$: Network-shared learnable penalty parameter.
\item $\widetilde{\mu}$: Network-shared learnable step size parameter.
\end{itemize}

\subsubsection{Adaptive Regularizer Module} 
This module, derived from Eq.~\ref{eq:nm z}, iteratively updates $\boldsymbol{Z}$. It employs a neural network to construct the nonlinear projection operator $\mathcal{F}_{\Theta}(\cdot)$ for regularization, as
\begin{equation}
    \boldsymbol{Z}^{(k)} = \,\mathcal{F}_{\Theta}(\boldsymbol{X}^{(k)} + \boldsymbol{V}^{(k-1)})
    \label{eq:adaptive regularization},
\end{equation}
where $\mathcal{F}_{\Theta}(\cdot)$ is a learnable regularizer detaied in Sec.~\ref{sec:ada_reg}.

\subsubsection{Scaled Multiplier Module}
This module updates $\boldsymbol{V}$ using a learnable, layer-shared rate $\widetilde{\eta}$, promoting stable convergence across the network. The computational process is formulated as
\begin{equation}
\boldsymbol{V}^{(k)} = \boldsymbol{V}^{(k-1)} + \widetilde{\eta}\left(\boldsymbol{X}^{(k)} - \boldsymbol{Z}^{(k)}\right),   
\end{equation}

ARSAR-Net eliminates the need for manual parameter tuning through automated optimization across three key modules. Unlike conventional deep networks, it derives its architecture from the ADMM unfolding framework, offering improved interpretability and theoretical grounding.

\subsection{Adaptive Regularizer}
\label{sec:ada_reg} 
Given complex-valued 2D matrices $\boldsymbol{X}$, $\boldsymbol{Z}$, and $\boldsymbol{V}$ in the network, ARSAR-Net incorporates specialized components for efficient complex data processing. 
The nonlinear projection operator $\mathcal{F}_{\Theta}$ in Eq.~\ref{eq:adaptive regularization} is implemented using convolutional layers for feature extraction, activation functions for nonlinear mapping, and normalization layers for training stability. 
Since these components are defined on real-valued data, we decompose each complex matrix into separate real and imaginary parts, expanding the data dimensions from $[B, H, W]$ to $[B, C, H, W]$, where $C$ denotes the number of channels. 
After processing, the real and imaginary parts are recombined to restore the original complex-valued structure.

To meet different application demands, we propose two ARSAR-Net variants: \textbf{Swift}, tailored for efficient inference, and \textbf{Pro}, optimized for high-quality reconstruction.

\subsubsection{ARSAR-Net Swift}
This module integrates a pyramid structure with multi-scale fusion blocks to reduce computational costs.
The feature pyramid network is employed as the backbone~\cite{lin2017feature}, which facilitates hierarchical feature extraction.
The multi-scale fusion block first applies interpolation-based upsampling to the down-sampled feature maps, then concatenates them with the same-scale feature maps along the channel dimension, which maintains spatial resolution while compressing channels.

\subsubsection{ARSAR-Net Pro}
This module employs a symmetric architecture that maintains consistent feature resolution across network layers, enhancing detail preservation at the cost of computational efficiency.
A convolutional cell consisting of 3×3 convolutional layers and activation functions is used and sequentially stacked to extract features. 
To ensure consistent channel dimensions between the input and output, a symmetric module is constructed to mirror this cell. 
In addition, skip-connections~\cite{he2016deep} are introduced outside the main structure to provide identity mappings, which facilitates preserving fine details and accelerating network convergence.

\subsection{Network Training}
\subsubsection{Loss function}
The training dataset consists of $N_{trn}$ samples, each defined as a pair $(Y_i, X_i)$, where $Y_i$ represents a complex-valued SAR echo signal and $X_i$ its corresponding ground truth image.
Conventional loss functions, such as Mean Squared Error (MSE), have been found unsuitable for complex-valued SAR image reconstruction. Through experiments, we empirically design the Normalized Mean Perceived Error (NMPE), a specialized loss function designed to be both compatible with complex signals and computationally efficient, as
\begin{equation}
\mathcal{L} = \frac{1}{N_{trn}}\sum_{i=1}^{N_{trn}} \frac{\frac{1}{H \times W}\sum_{p=1}^H\sum_{q=1}^W(|\hat{X}_i(p,q)| - |X_i(p,q)|)^2}{\lVert X_i \rVert_2}
\end{equation}
where $\hat{X}_i$ and $Y_i$ respectively denote ARSAR-Net's output and input, $\lVert \cdot \rVert_2$ denotes the $\ell_2$-norm.

\subsubsection{Training Configuration}
We implement ARSAR-Net in the PyTorch framework, using the Adam optimizer~\cite{kingma2014adam} with a learning rate of 2e-5 across 80 training epochs. Both variants of ARSAR-Net share identical configurations: a 9-layer architecture and a batch size of 4 during training.

\section{Experiment}
\label{sec:exper}

\subsection{Experimental Setting}
\label{sec:exper_conf}
\subsubsection{SAR System Parameters}
All the real SAR data used in the experiment are acquired from GaoFen-3, a C-band polarimetric SAR satellite, which works in a vertical side-looking strip mode (Fig.~\ref{radar}). 
Key parameters of this system are given in table~\ref{tab:sar_params}, which are also adopted in the generation of the simulated echo data.

\begin{table}[H]
\centering
\caption{Main Parameters of the SAR System}
\label{tab:sar_params}
\begin{tabular}{cc}
\toprule
\textbf{Parameter}         & \textbf{Value} \\ 
\midrule
Signal bandwidth           & 60 MHz \\ 
Pulse width                & 45 us  \\ 
Pulse repetition frequency & 1420 Hz \\ 
Carrier frequency          & 5.4 GHz \\ 
Slant range                & 850 km \\ 
Equivalent velocity        & 7500 m/s \\  
\bottomrule
\end{tabular}
\end{table}

\begin{table}[H]
\centering
\caption{Dataset Configuration}
\label{tab:dataset}
\begin{tabular}{ccc}
\toprule
\textbf{Label} & \textbf{Training} & \textbf{Testing} \\ 
\midrule
Off-shore Ships & 800 & 200 \\ 
Harbors         & 400 & 100 \\ 
Cities          & 400 & 100 \\  
\bottomrule
\end{tabular}
\end{table}

\subsubsection{Experimental Platform}
All the experiments are conducted on a platform with an Intel Xeon Gold 5320 CPU (2.20 GHz) and NVIDIA A100-80GB GPU.

\subsubsection{Dataset}

For network training, we construct a dataset from SAR SLC images with 1m resolution acquired by the Gaofen-3 satellite, covering three scene types: offshore ships, harbors, and cities. The dataset includes 2,000 images standardized to $512 \times 512$ pixels, with 1,600 used for training and 400 for testing (Table~\ref{tab:dataset}).

Simulated echo signals are generated by applying the CSA inverse imaging operator (parameterized as in Table~\ref{tab:sar_params}) to the SLC images, which serve as ground truth. 
Due to system constraints, range sampling is limited by bandwidth, while azimuth sampling regulated by the pulse repetition frequency (PRF) allows more flexibility~\cite{moreira1993suppressing}, making azimuth downsampling effective for practical use.
We adopt a random downsampling matrix $\mathbf{\Psi}$, which satisfies RIP conditions~\cite{baraniuk2008simple,castorena2014restricted}, to create azimuth-downsampled echoes at 50\% and 75\% rates for training and evaluation.

\subsubsection{Evaluation Metrics}
To quantitatively evaluate ARSAR-Net, the following metrics are employed: NRMSE~\cite{poli1993use}, Peak Signal-to-Noise Ratio (PSNR, measured in dB), and Structural Similarity Index for Measuring (SSIM)~\cite{hore2010image}.

\subsection{Distributed Target Simulation}
\label{sec:exper_sim}

We conduct two simulation experiments to evaluate ARSAR-Net 
and compare it with sparse image methods CSA~\cite{dong2013novel}, $\ell_1$-Norm~\cite{candes2006robust}, TV regularization~\cite{chambolle2004algorithm}, and unfolding-based methods LRSR-ADMM-Net~\cite{an2022lrsr} and SR-CSA-Net~\cite{zhang2023nonsparse}.
In these experiments, the simulated echoes are azimuth-downsampled at two rates (50\% and 75\%) to assess imaging performance under varying sampling conditions, with quantitative results summarized in Table~\ref{tab:simulation}. 
For a fair comparison, all methods use identical downsampled echo data under consistent experimental settings.

\begin{table*}[t]
\caption{Comparison of Imaging Performance under Sampling Rates for Simulated Data}
\label{tab:simulation}
\centering
\resizebox{0.7\textwidth}{!}{
\begin{tabular}{lcccccc}
\toprule
\multirow{2}{*}{Method}  &\multicolumn{3}{c}{50\% Sampling Rate} &\multicolumn{3}{c}{75\% Sampling Rate} \\ 
\cmidrule(r){2-4} \cmidrule(r){5-7}
& NRMSE & PSNR  & SSIM & NRMSE & PSNR  & SSIM \\
\midrule
CSA~\cite{dong2013novel} & 0.605 & 12.52 & 0.424 & 0.460 & 14.99 & 0.485 \\
$\ell_1$-Norm~\cite{candes2006robust} & 0.521 & 14.16 & 0.574 & 0.203 & 20.04 & 0.791 \\
TV~\cite{chambolle2004algorithm} & 0.544 & 13.37 & 0.512 & 0.374 & 16.53 & 0.608 \\
\midrule
LRSR-ADMM-Net~\cite{an2022lrsr} & 0.722 & 11.32 & 0.375 & 0.418 & 15.61 & 0.491 \\
SR-CSA-Net~\cite{zhang2023nonsparse} & 0.209 & 20.24 & 0.683 & 0.159 & 22.78 & 0.861 \\
\midrule
ARSAR-Net Swift & 0.239 & 18.83 & 0.685 & 0.086 & 28.17 & 0.861 \\
ARSAR-Net Pro & \textbf{0.140} & \textbf{24.54} & \textbf{0.862} & \textbf{0.047} & \textbf{32.41} & \textbf{0.948} \\
\bottomrule
\end{tabular}
}
\end{table*}

\begin{figure*}[t]
    \centering
    \subfloat[]{\includegraphics[width=0.18\textwidth]{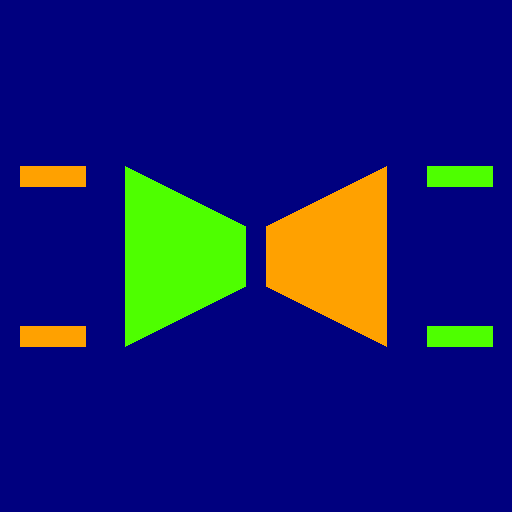}\label{fig:sim-50-a}}\hfill
    \subfloat[]{\includegraphics[width=0.18\textwidth]{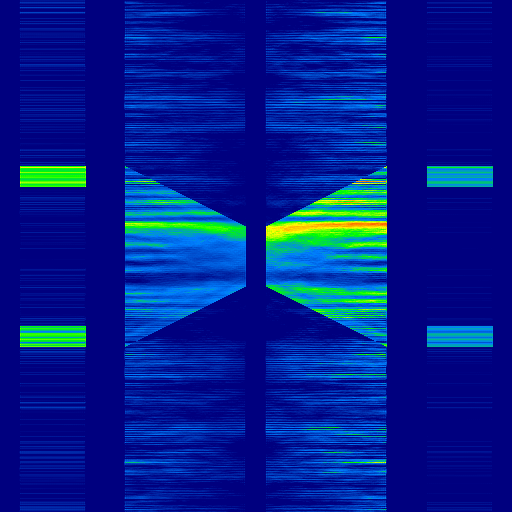}\label{fig:sim-50-b}}\hfill
    \subfloat[]{\includegraphics[width=0.18\textwidth]{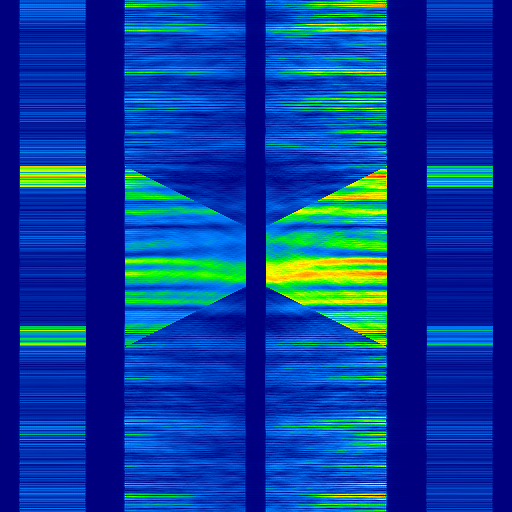}\label{fig:sim-50-c}}\hfill
    \subfloat[]{\includegraphics[width=0.18\textwidth]{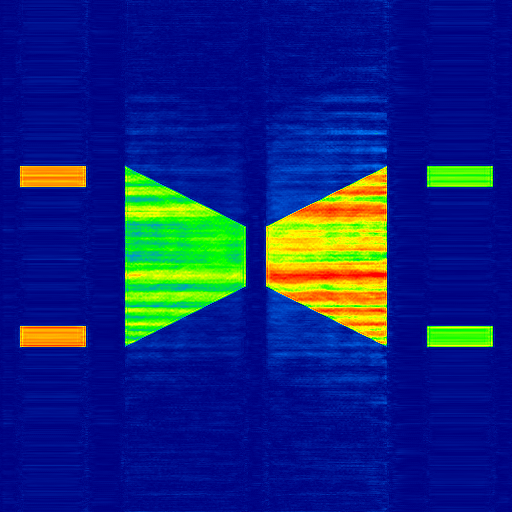}\label{fig:sim-50-d}}\hfill
    \subfloat[]{\includegraphics[width=0.18\textwidth]{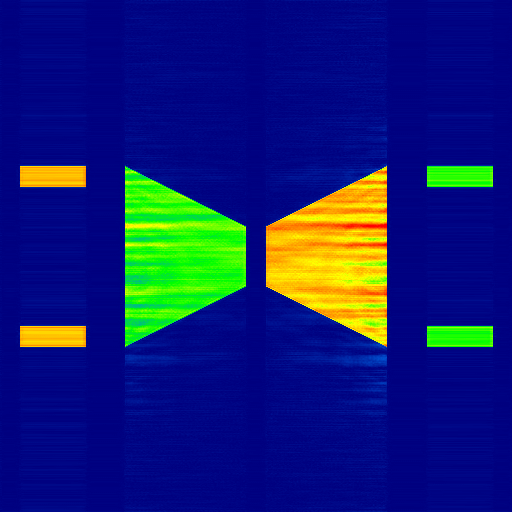}\label{fig:sim-50-e}}

    \caption{Comparison of image reconstruction results under the 50\% sampling rate. (a) a simulated image. (b) $\ell_1$-Norm, (c) LRSR-ADMM-Net, (d) SR-CSA-Net, and (e) ARSAR-Net Pro.}
    \label{sim-50}
\end{figure*}

With a 50\% downsampling rate, conventional methods including CSA $\ell_1$-Norm and TV regularization (Fig.~\ref{fig:sim-50-b}) demonstrate suboptimal imaging performance, which implies poor PSNR and SSIM and NRMSE values.
LRSR-ADMM-Net (Fig.~\ref{fig:sim-50-c}) performs poorly, while SR-CSA-Net (Fig.~\ref{fig:sim-50-d}) offers moderate improvements by partially suppressing ambiguities and enhancing edges.
ARSAR-Net (Fig.~\ref{fig:sim-50-e}) further improves reconstruction quality by reducing azimuth ambiguity and restoring structural details. The Pro variant achieves the best results, with a PSNR of 24.54 dB and SSIM of 0.862, Table ~\ref{tab:simulation}.

At 75\% downsampling, all methods show improved performance. Conventional approaches still suffer from detail loss and azimuth ambiguity, while SR-CSA-Net achieves better edge preservation, reaching a PSNR of 22.78 dB and an SSIM of 0.861, Table ~\ref{tab:simulation}. 
ARSAR-Net Swift provides further enhancements, and the Pro variant attains the highest quality, with a PSNR of 32.41 dB and an SSIM of 0.948, preserving fine textures and overall structural consistency.
These demonstrate strong generalization and high image fidelity.

\subsection{Imaging in Real Scenes}
\label{sec:exper_real}

\subsubsection{Overall dataset}
The first experiment is carried out on the overall dataset under 50\% and 75\% sampling rates, comparing its performance against both conventional (CSA, $\ell_1$-norm and TV)  and unfolding networks.
Unfolding networks like LRSR-ADMM-Net\footnote{While LRSR-ADMM-Net demonstrated promising results~\cite{an2022lrsr}, our reproduction reveals a noticeable performance gap in real-scene reconstruction, which we attribute to dataset variations between the two studies.} and SR-CSA-Net for a more comprehensive comparison.
The overall dataset contains a diverse mix of real-world SAR scenes, covering scenes of different sparsity levels. Experimental results are included in Table~\ref{tab:overall comparison}.

\begin{table*}[t]
\caption{Comparison of Imaging Performance under Sampling Rates for Real Scenes}
\label{tab:overall comparison}
\centering
\resizebox{0.9\textwidth}{!}{
\begin{tabular}{lccccccc}
\toprule
\multirow{2}{*}{Method}  &\multicolumn{3}{c}{50\% Sampling Rate} &\multicolumn{3}{c}{75\% Sampling Rate} & \multirow{2}{*}{Imaging Speed (items/s)} \\ 
\cmidrule(r){2-4} \cmidrule(r){5-7}
& NRMSE & PSNR & SSIM & NRMSE & PSNR & SSIM \\
\midrule
CSA & 0.384 & 24.02 & 0.539 & 0.287 & 26.78 & 0.693 & 210.2\\
$\ell_1$-Norm & 0.675 & 22.79 & 0.286 & 0.676 & 24.01 & 0.318 & 3.656 \\
TV & 0.377 & 24.07 & 0.550 & 0.276 & 26.36 & 0.694 & 3.371 \\
\midrule
LRSR-ADMM-Net & 0.349 & 24.11 & 0.543 & 0.186 & 27.00 & 0.688 & 1.947\\
SR-CSA-Net & 0.162 & 28.91 & 0.756 & 0.077 & 34.17 & 0.903 & 43.84 \\
\midrule
ARSAR-Net Swift & 0.089 & 29.15 & 0.743 & 0.062 & 33.29 & 0.881 & \textbf{67.60} \\
ARSAR-Net Pro & \textbf{0.084} & \textbf{30.75} & \textbf{0.810} & \textbf{0.036} & \textbf{36.05} & \textbf{0.928} & 17.60 \\
\bottomrule
\end{tabular}
}
\end{table*}





\begin{figure*}[t]
    \centering

    \subfloat[]{\includegraphics[width=0.18\textwidth]{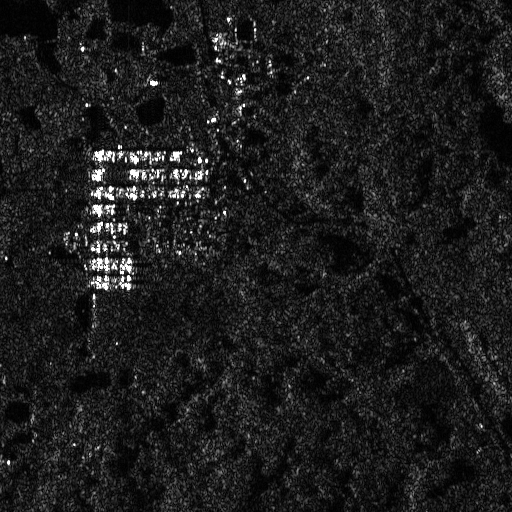} \label{fig:real-sparse-50-a}}\hfill
    \subfloat[]{\includegraphics[width=0.18\textwidth]{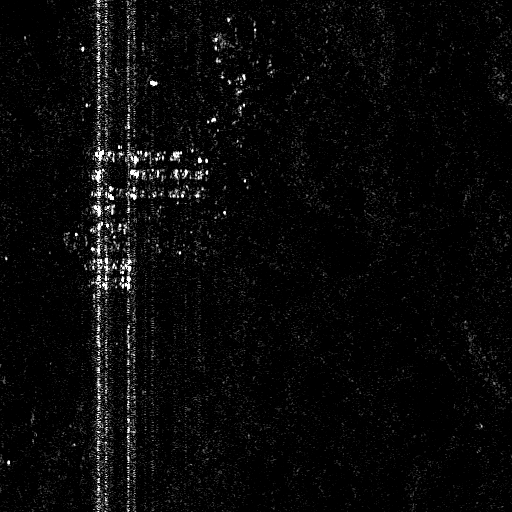} \label{fig:real-sparse-50-b}}\hfill
    \subfloat[]{\includegraphics[width=0.18\textwidth]{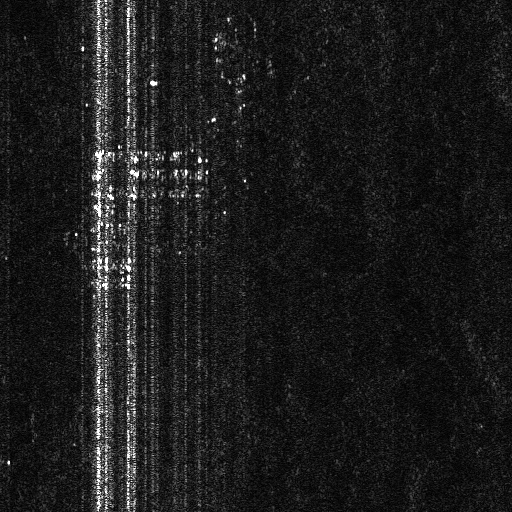} \label{fig:real-sparse-50-c}}\hfill
    \subfloat[]{\includegraphics[width=0.18\textwidth]{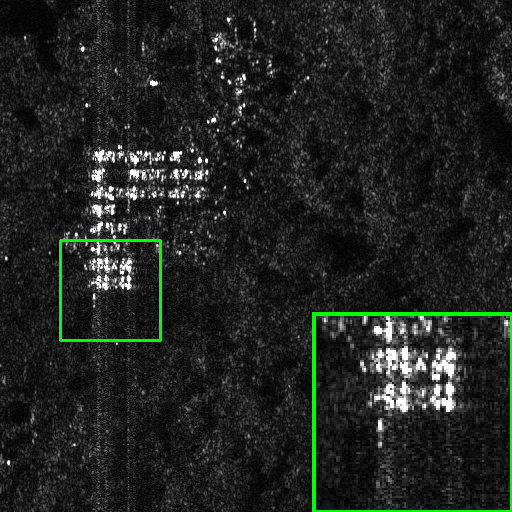} \label{fig:real-sparse-50-d}}\hfill
    \subfloat[]{\includegraphics[width=0.18\textwidth]{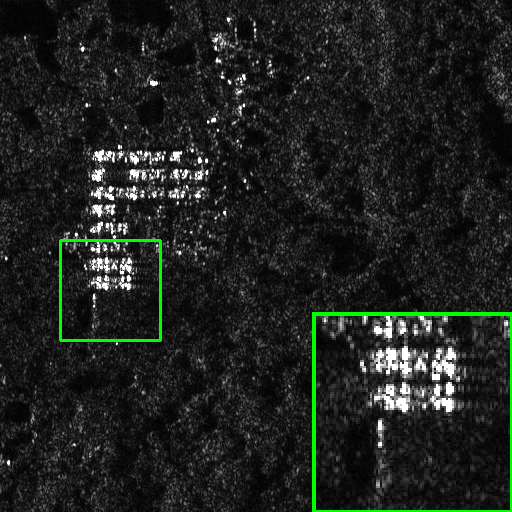} \label{fig:real-sparse-50-e}}

    \caption{Comparison of imaging results in a real sparse scene under a 50\% sampling rate. (a) ground truth. (b) $\ell_1$-Norm. (c) LRSR-ADMM-Net. (d) SR-CSA-Net. (e) ARSAR-Net Pro.}
    \label{fig:real-sparse-50}
\end{figure*}

\begin{figure*}[t]
    \centering

    \subfloat[]{\includegraphics[width=0.18\textwidth]{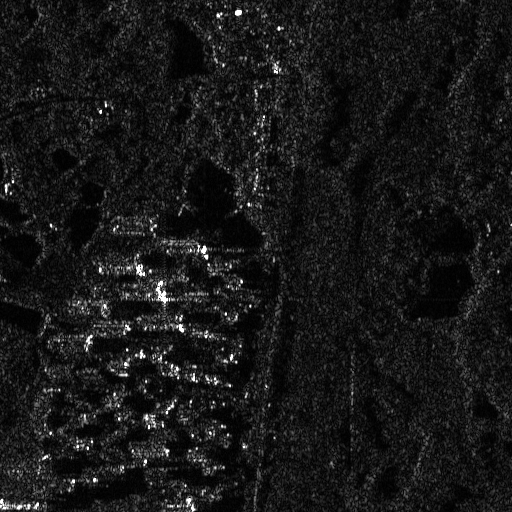} \label{fig:real-nonsparse-50-a}}\hfill
    \subfloat[]{\includegraphics[width=0.18\textwidth]{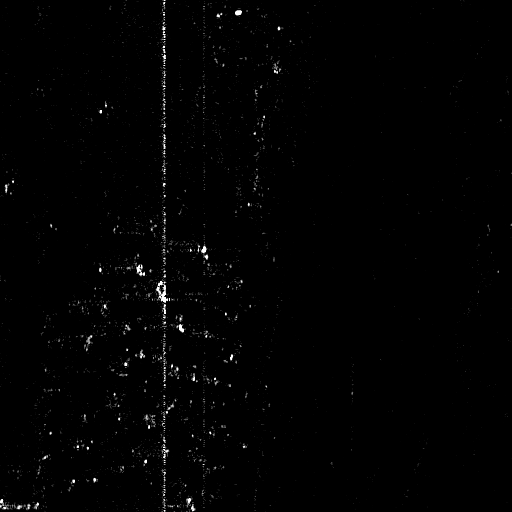} \label{fig:real-nonsparse-50-b}}\hfill
    \subfloat[]{\includegraphics[width=0.18\textwidth]{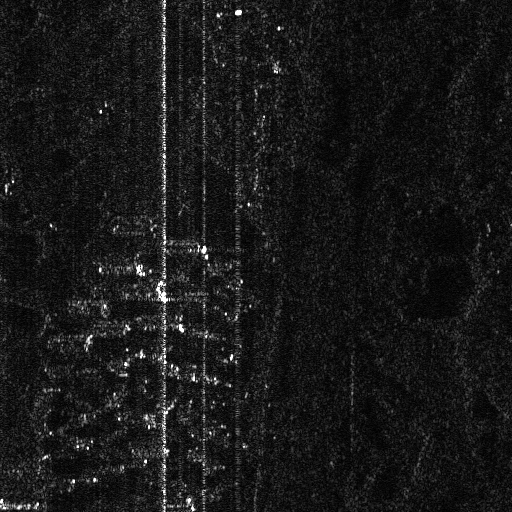} \label{fig:real-nonsparse-50-c}}\hfill
    \subfloat[]{\includegraphics[width=0.18\textwidth]{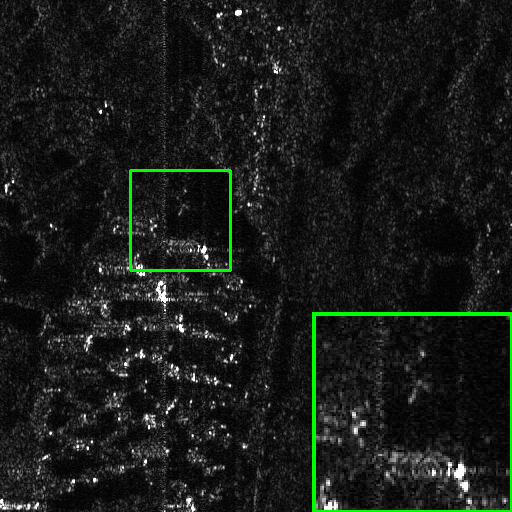} \label{fig:real-nonsparse-50-d}}\hfill
    \subfloat[]{\includegraphics[width=0.18\textwidth]{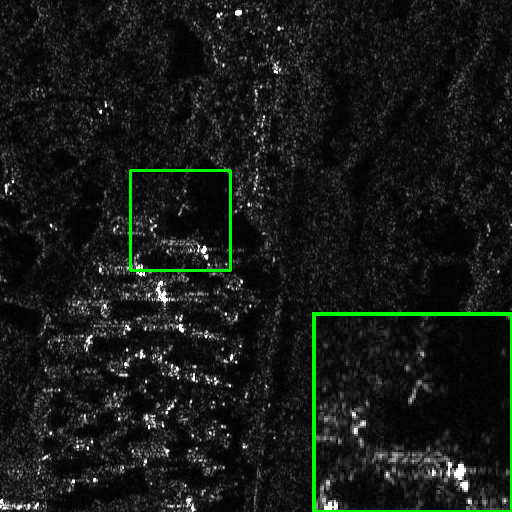} \label{fig:real-nonsparse-50-e}}

    \caption{Comparison of imaging results in a real non-sparse scene under a 50\% sampling rate. (a) ground truth. (b) $\ell_1$-Norm. (c) LRSR-ADMM-Net. (d) SR-CSA-Net. (e) ARSAR-Net Pro.}
    \label{fig:real-nonsparse-50}
\end{figure*}

In Fig.~\ref{fig:real-sparse-50}, reconstruction results suffer from the azimuth ambiguity under the 50\% sampling rate. CSA-based imaging offers limited suppression, while TV regularization reduces noise-induced gradients but struggles with severe ambiguities. 
The $\ell_1$-norm method (Fig.~\ref{fig:real-sparse-50-b}) better suppresses ambiguities in sparse scenes but sacrifices fine details in non-sparse scenes.
Unfolding networks such as LRSR-ADMM-Net and SR-CSA-Net (Figs.~\ref{fig:real-sparse-50-c} and~\ref{fig:real-sparse-50-d}) show moderate improvements, balancing ambiguity suppression and detail preservation. 
ARSAR-Net Swift achieves higher quality across scenes, though its downsampling leads to some detail loss. ARSAR-Net Pro (Fig.~\ref{fig:real-sparse-50-e}) outperforms all baselines, reaching 30.75 dB PSNR and 0.810 SSIM.
At 75\% sampling, azimuth ambiguity is further reduced. Conventional methods improve modestly but still suffer from residual artifacts. SR-CSA-Net, ARSAR-Net Swift, and Pro show clear advantages, with ARSAR-Net Pro achieving the best results (36.05 dB PSNR, 0.928 SSIM), maintaining fine textures and strong structural consistency.

Imaging efficiency is crucial for practical applications.
CSA-based operators achieve the highest speed of 210.2 items/s but at the cost of image quality.
Conventional approaches are slower due to their iterative optimization procedures requiring repeated updates.
The unfolding network, LRSR-ADMM-Net, is bottlenecked by the singular value decomposition operations, whereas SR-CSA-Net benefits from a more efficient unfolding design.
Our ARSAR-Net Swift improves efficiency with a pyramid structure, reaching a superior imaging speed of 67.60 items/s. In contrast, ARSAR-Net Pro emphasizes reconstruction quality with a full-resolution design, trading speed for higher precision.

\begin{figure*}[t]
    \centering

    \subfloat[]{\includegraphics[width=0.18\textwidth]{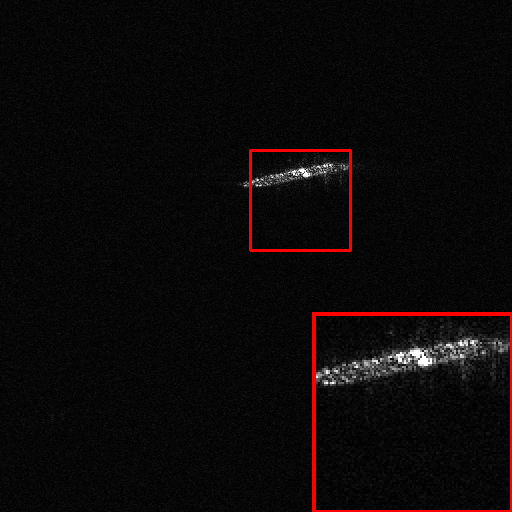}} \hfill
    \subfloat[]{\includegraphics[width=0.18\textwidth]{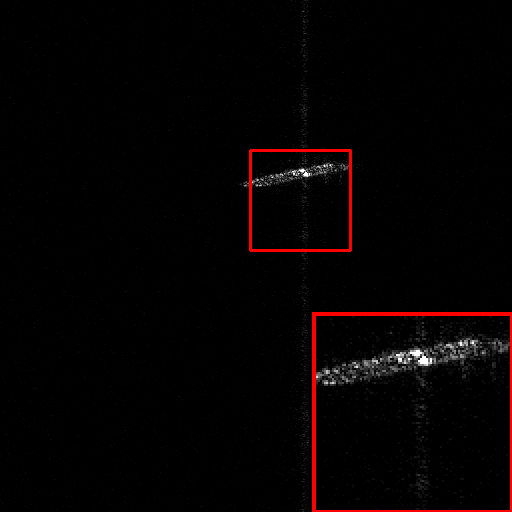}} \hfill
    \subfloat[]{\includegraphics[width=0.18\textwidth]{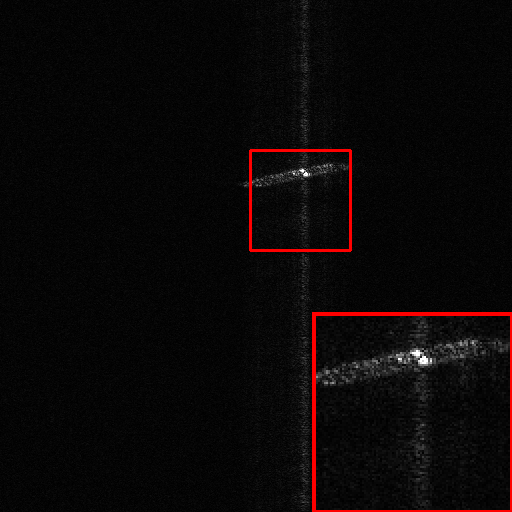}} \hfill
    \subfloat[]{\includegraphics[width=0.18\textwidth]{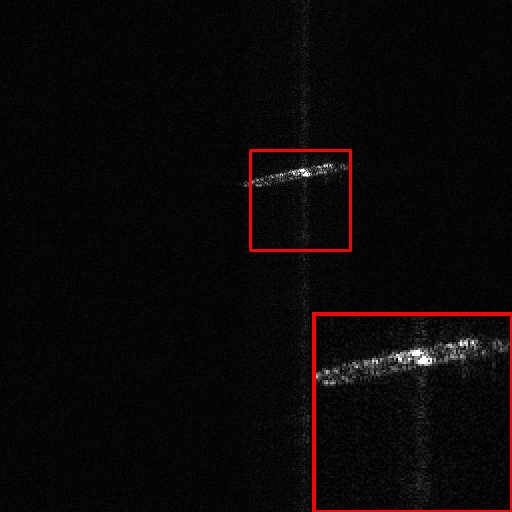}} \hfill
    \subfloat[]{\includegraphics[width=0.18\textwidth]{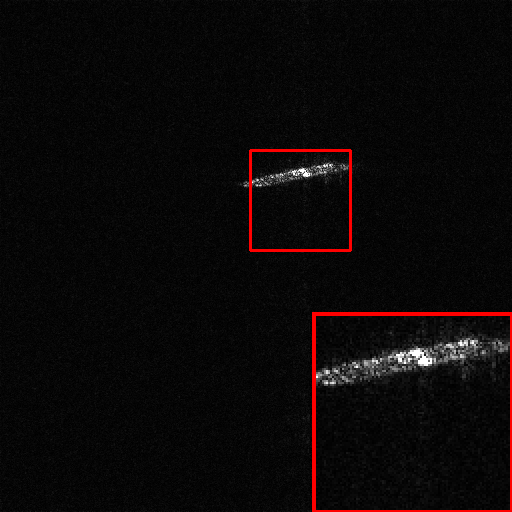}}

    \caption{Comparison of imaging results in a sparse scene. (a) ground truth. (b) $\ell_1$-Norm. (c) LRSR-ADMM-Net. (d) SR-CSA-Net. (e) ARSAR-Net.}
    \label{fig:subset-sparse}
\end{figure*}

\begin{figure*}[t]
    \centering

    \subfloat[]{\includegraphics[width=0.18\textwidth]{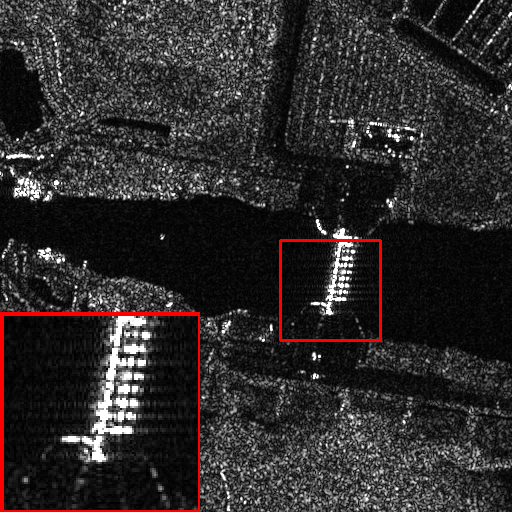}} \hfill
    \subfloat[]{\includegraphics[width=0.18\textwidth]{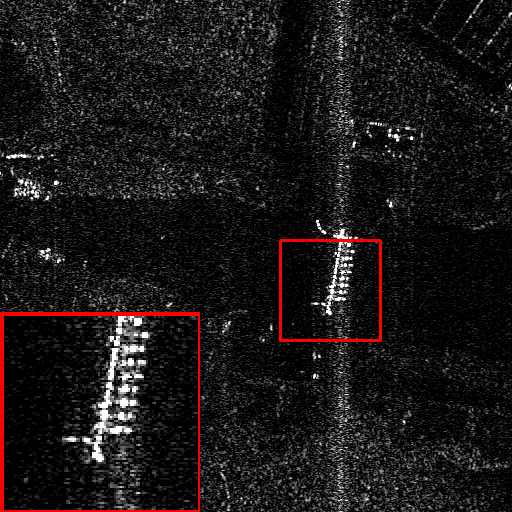}} \hfill
    \subfloat[]{\includegraphics[width=0.18\textwidth]{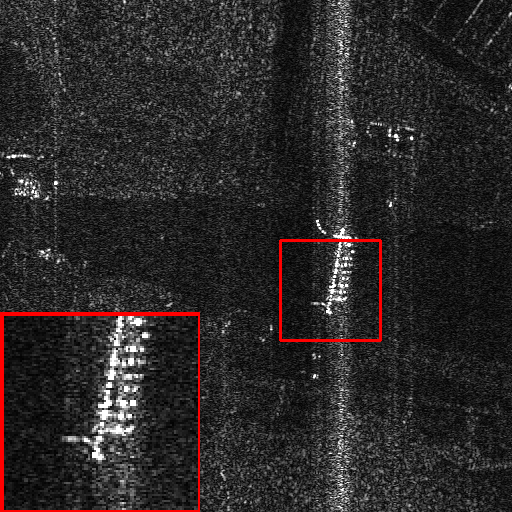}} \hfill
    \subfloat[]{\includegraphics[width=0.18\textwidth]{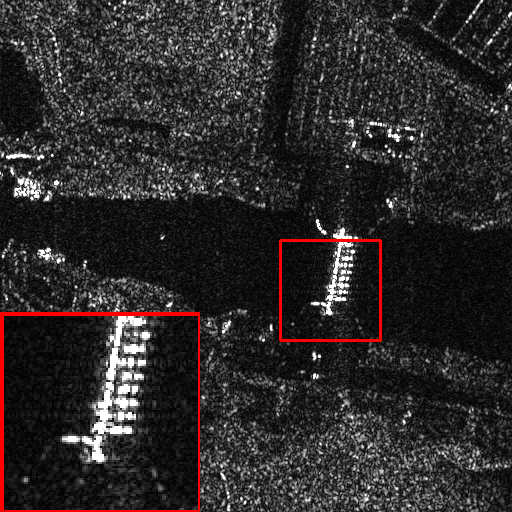}} \hfill
    \subfloat[]{\includegraphics[width=0.18\textwidth]{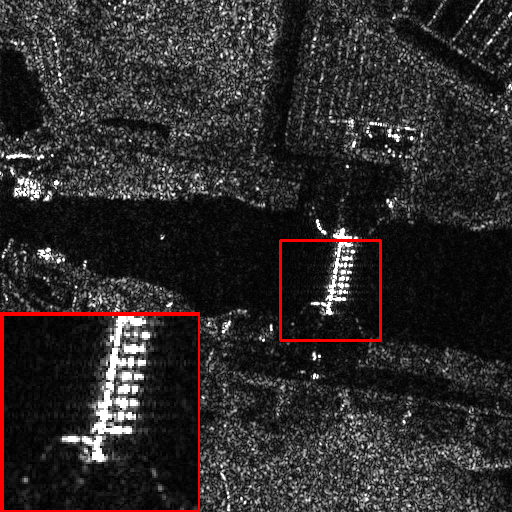}}

    \caption{Comparison of imaging results in a non-sparse scene. (a) ground truth. (b) $\ell_1$-Norm. (c) LRSR-ADMM-Net. (d) SR-CSA-Net. (e) ARSAR-Net.}
    \label{fig:subset-non-sparse}
\end{figure*}

\subsubsection{Sparse and non-sparse scenes}

This experiment assesses adaptability in scenes of different sparsity levels. 
The compared approaches consist of a conventional method based on $\ell_1$-norm and two unfolding networks: LRSR-ADMM-Net and SR-CSA-Net. For a fair comparison, all data-driven methods are trained on the overall dataset. 

\begin{table*}[t]
\caption{Comparison of Imaging Performance under Sampling Rates in Subsets}
\label{tab:subset}
\centering
\resizebox{0.8\textwidth}{!}{
\begin{tabular}{cccccccc}
\toprule
\multirow{2}{*}{Scene Type} & \multirow{2}{*}{Method} & \multicolumn{3}{c}{50\% Sampling Rate} & \multicolumn{3}{c}{75\% Sampling Rate} \\ 
\cmidrule(r){3-5} \cmidrule(r){6-8}
 & & NRMSE & PSNR & SSIM & NRMSE & PSNR & SSIM \\
\midrule
\multirow{4}{*}{Sparse} 
& $\ell_1$-Norm & 0.747 & 29.21 & 0.319 & 0.752 & 29.91 & 0.330 \\
& LRSR-ADMM-Net & 0.334 & 31.23 & 0.746 & 0.173 & 34.27 & 0.853 \\
& SR-CSA-Net & 0.278 & 32.91 & 0.777 & 0.147 & 35.98 & 0.877 \\
& ARSAR-Net & \textbf{0.134} & \textbf{34.60} & \textbf{0.802} & \textbf{0.056} & \textbf{37.93} & \textbf{0.902} \\
\midrule
\multirow{4}{*}{Non-sparse}
& $\ell_1$-Norm & 0.615 & 18.24 & 0.277 & 0.611 & 19.93 & 0.328 \\
& LRSR-ADMM-Net & 0.355 & 19.07 & 0.397 & 0.191 & 21.89 & 0.567 \\
& SR-CSA-Net & 0.086 & 26.00 & 0.743 & 0.031 & 32.78 & 0.921 \\
& ARSAR-Net & \textbf{0.050} & \textbf{28.01} & \textbf{0.817} & \textbf{0.023} & \textbf{34.65} & \textbf{0.947} \\
\bottomrule
\end{tabular}
}
\end{table*}

For the case of the 50\% sampling rate, Fig.~\ref{fig:subset-sparse} and Fig.~\ref{fig:subset-non-sparse} show the imaging results in both sparse and non-sparse scenes.
The $\ell_1$-norm method effectively suppresses noise and ambiguity but suffers from severe detail loss. In sparse scenes, its PSNR and SSIM are lower by approximately 5.5 dB and 0.5 compared to unfolding methods, Table~\ref{tab:subset}. 
In non-sparse scenes, the gap widens to 10 dB and 0.55, highlighting its limitations for SAR imaging under downsampled conditions.

LRSR-ADMM-Net achieves promising imaging quality in sparse scenes, with PSNR and SSIM values comparable to the best-performing method.
Nevertheless, due to its limited parameter size and training capacity, its performance degrades in non-sparse scenes, $e.g.$, PSNR dropping by about 9 dB and SSIM decreasing by 0.4 compared to ARSAR-Net. 
This substantial performance gap indicates that LRSR-ADMM-Net lacks the adaptability to non-sparse scenes.
By integrating an unfolding network with the CSA-based imaging operators, SR-CSA-Net achieves high imaging performance in both sparse and non-sparse scenes.
The proposed ARSAR-Net further improves reconstruction fidelity through the adaptive regularizer module, outperforming existing methods by 2 dB PSNR. 
Moreover, as shown in Table~\ref{tab:scene-agnostic}, ARSAR-Net achieves the highest PSNR across all sparsity levels, further validating its scene-agnostic imaging capability.

\begin{table}[H]
\centering
\caption{PSNR (dB) Comparison of Different Methods under 50\% Sampling Rate Across Varying Scene Sparsity}
\label{tab:scene-agnostic}
\begin{tabular}{lcccc}
\toprule
\textbf{Method} & \textbf{Very Sparse} & \textbf{Sparse} & \textbf{Normal} & \textbf{Non-sparse}  \\ 
\midrule
LRSR-ADMM-Net     &  32.32          &  29.65          &  23.53          &  14.29           \\ 
SR-CSA-Net        &  33.58          &  32.47          &  28.22          &  24.10           \\
ARSAR-Net         &  \textbf{34.91} &  \textbf{33.83} &  \textbf{30.71} &  \textbf{25.69}  \\
\bottomrule
\end{tabular}
\end{table}

\FloatBarrier

\begin{table*}[t]
\caption{Performance of ARSAR-Net Variants under Different Layer Numbers under 50\% Downsampling Rate}
\label{tab:layer-number}
\centering
\resizebox{0.8\textwidth}{!}{
\begin{tabular}{lcccccccccc}
\toprule
\multirow{2}{*}{Method}  &\multicolumn{3}{c}{$N_s=3$} &\multicolumn{3}{c}{$N_s=6$} &\multicolumn{3}{c}{$N_s=9$}\\ 
\cmidrule(r){2-4} \cmidrule(r){5-7} \cmidrule(r){8-10}
& PSNR & SSIM & Speed & PSNR & SSIM & Speed & PSNR & SSIM & Speed \\
\midrule
ARSAR-Net Swift & 28.64 & 0.728 & 154.0 & 28.99 & 0.739 & 71.23 & 29.15 & 0.743 & 67.60 \\
ARSAR-Net Pro & 24.03 & 0.540 & 33.28 & 29.20 & 0.783 & 21.92 & 30.75 & 0.810 & 17.60 \\
\bottomrule
\end{tabular}
}
\end{table*}

\begin{figure*}[t]
    \centering
    \captionsetup[subfigure]{labelformat=empty}

    \subfloat[1st Layer(10.87dB)]{\includegraphics[width=0.18\textwidth]{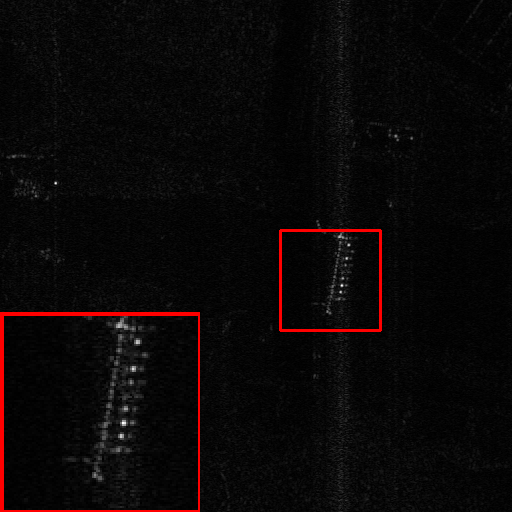}} \hfill
    \subfloat[3rd Layer(24.03dB)]{\includegraphics[width=0.18\textwidth]{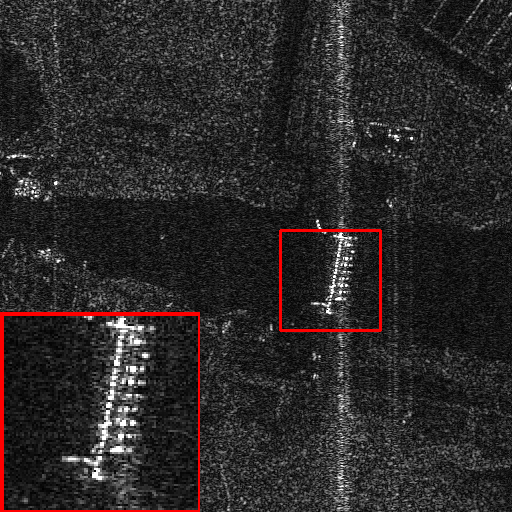}} \hfill
    \subfloat[5th Layer(27.24dB)]{\includegraphics[width=0.18\textwidth]{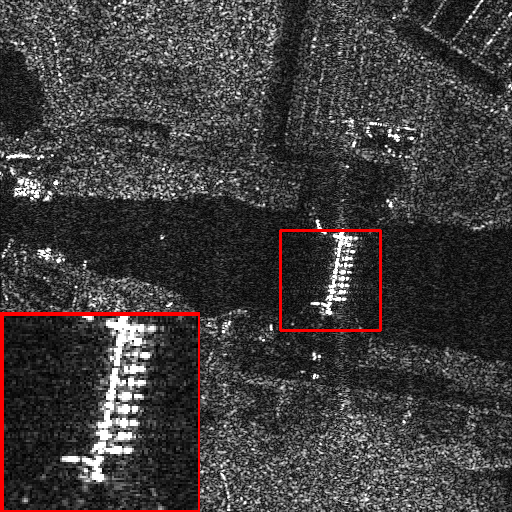}} \hfill
    \subfloat[7th Layer(29.80dB)]{\includegraphics[width=0.18\textwidth]{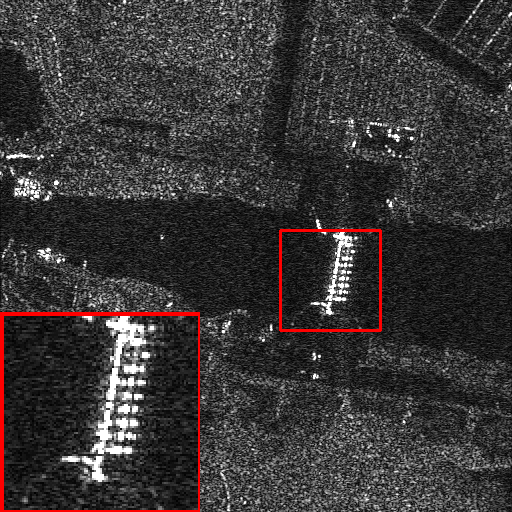}} \hfill
    \subfloat[9th Layer(30.75dB)]{\includegraphics[width=0.18\textwidth]{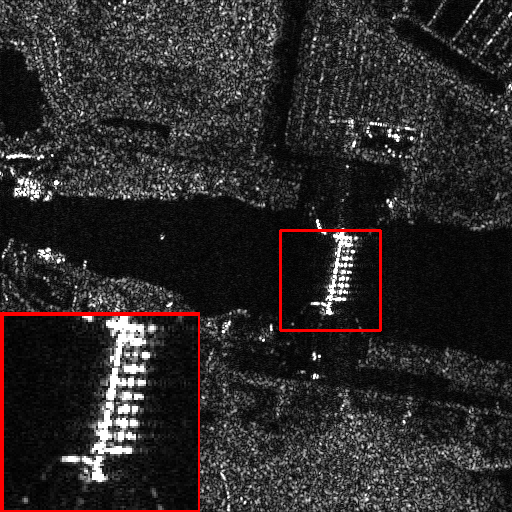}}
    \caption{Reconstruction results of ARSAR-Net Pro at different layers under 50\% sampling rate, with PSNR values shown.}
    \label{fig:arsar-layerwise}
\end{figure*}

\subsubsection{Unfolding network layer setting}
ARSAR-Net is derived by unfolding the ADMM algorithm without matrix inversion, where each layer corresponds to an iteration.
Therefore, the network layer number ($N_s$) has a significant impact on both imaging speed (items/s) and reconstruction quality, Table~\ref{tab:layer-number}. As shown in Fig.~\ref{fig:arsar-layerwise}, the outputs of ARSAR-Net Pro from layers 1, 3, 5, 7, and 9 demonstrate a progressive improvement in reconstruction quality under a 50\% sampling rate.

\begin{figure*}[t]
\begin{minipage}[b]{0.54\textwidth}
    \centering
    \includegraphics[width=\textwidth]{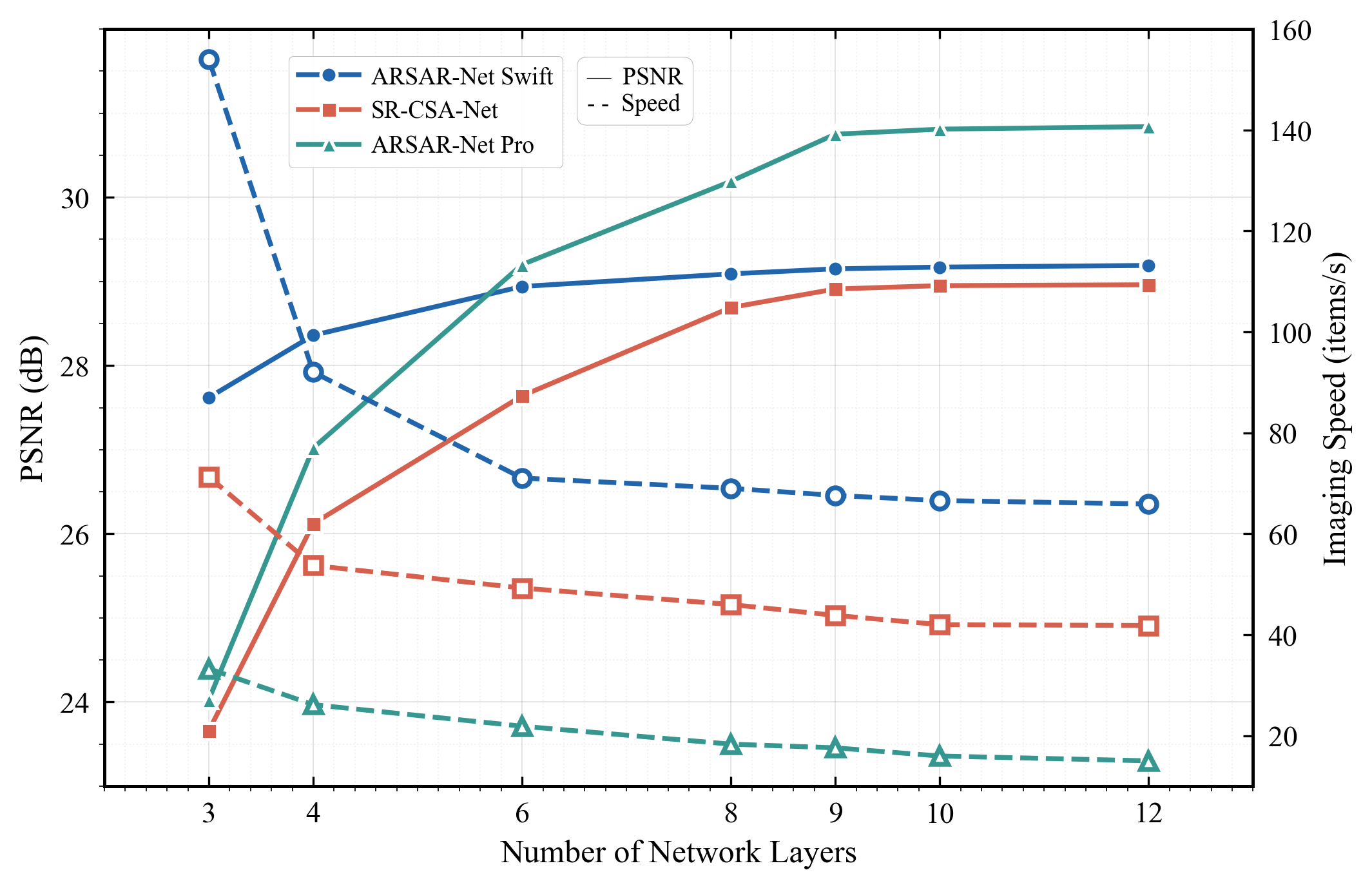}
    \caption{Performance comparison under different network layers.}
    \label{fig:compare_fig}
\end{minipage}
\hfill
\begin{minipage}[b]{0.46\textwidth}
    \centering
    \includegraphics[width=\textwidth]{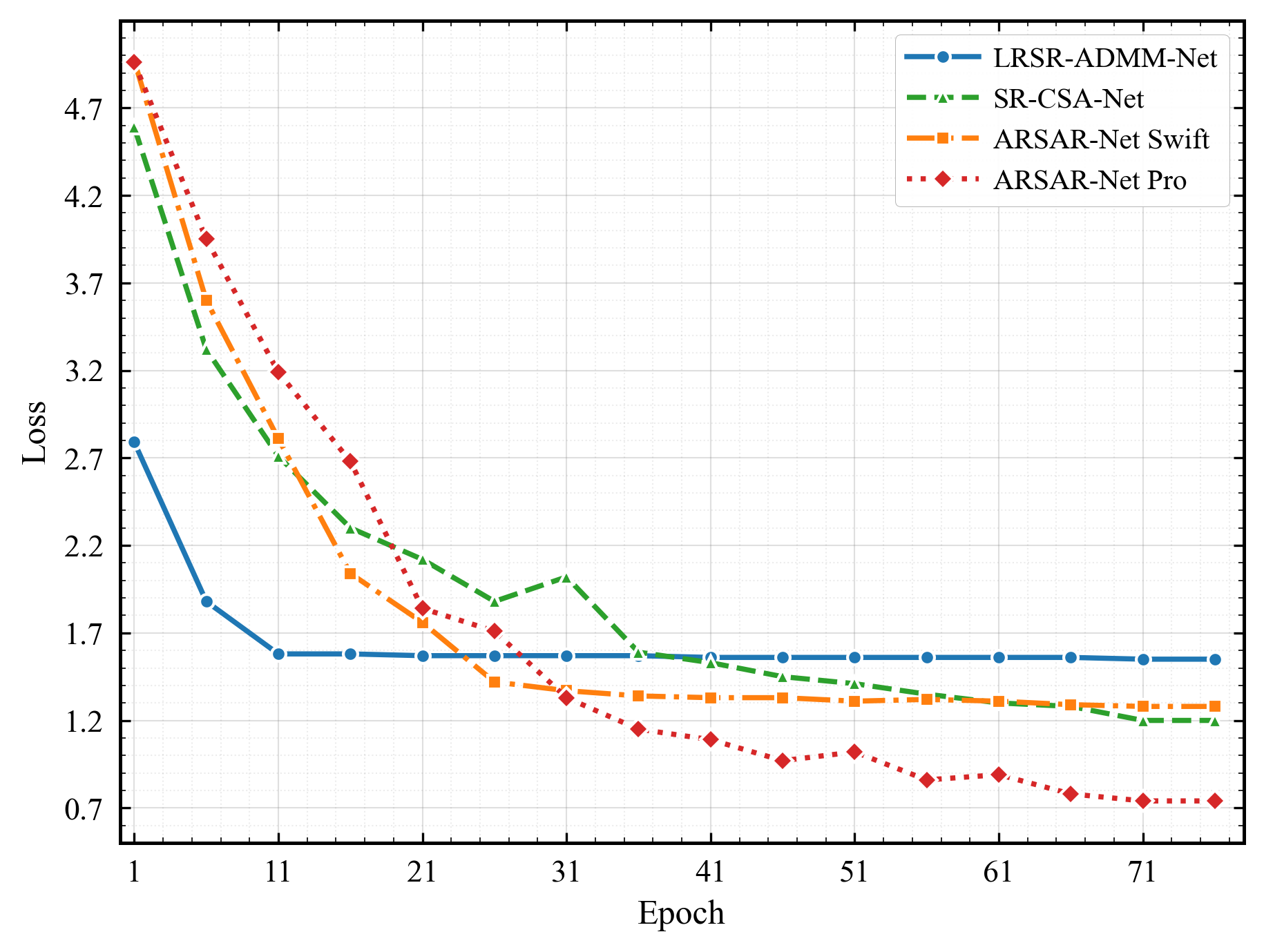}
    \caption{Training loss curves of compared methods}
    \label{fig:loss}
\end{minipage}
\end{figure*}

In Fig.~\ref{fig:compare_fig}, the PSNR value of ARSAR-Net Swift increases gradually with network layers and begins to plateau around 29.2 dB when the depth exceeds 9. In contrast, ARSAR-Net Pro shows a sharp improvement in PSNR as the depth increases, eventually stabilizing at about 30.8 dB beyond 8 layers. On the other hand, the imaging speed of both variants decreases rapidly as the number of layers increases.

For the trade-off between quality and speed, we select a 9-layer network for both ARSAR-Net variants. This shows superior imaging performance compared to networks with fewer layers while maintaining high imaging efficiency. 

\subsection{Training Convergence}
\label{sec:exper_convergence}

In Fig.~\ref{fig:loss}, LRSR-ADMM-Net shows a relatively flat, stable loss curve, suggesting limited adaptability due to its small parameter set. In contrast, SR-CSA-Net converges rapidly, with loss dropping from over 4.6 to 1.2, indicating effective training dynamics but raising concerns about potential generalization limitations.
ARSAR-Net Swift begins with the highest initial loss but decreases steadily, indicating a stable training process with consistent improvements. Although its final loss is higher than ARSAR-Net Pro, the trend remains reliable. ARSAR-Net Pro achieves the lowest loss (around 0.7), reflecting efficient optimization and superior convergence.
Overall, ARSAR-Net Pro enjoys the best convergence performance, achieving a balance between speed and final loss. In comparison, ARSAR-Net Swift converges more gradually, reflecting a stable and robust optimization process.

\section{Conclusion}
We proposed ARSAR-Net, an efficient unfolding network with adaptive regularization towards scene-agnostic SAR imaging.
By integrating a learnable regularization module into an unfolding architecture, ARSAR-Net enables adaptive and data-driven reconstruction across diverse environments.
Built upon the efficient ADMM without matrix inversion, ARSAR-Net enjoys computational efficiency.
To further meet application-specific needs, two variants of ARSAR-Net were introduced: ARSAR-Net Swift for accelerated imaging and ARSAR-Net Pro for high-fidelity reconstruction.
Extensive experiments on both simulated and real SAR datasets validate the effectiveness of our approach, demonstrating significant gains in speed, image quality, and scene adaptability.
%
ARSAR-Net provides a fresh insight into high-quality and efficient SAR imaging in complex scenes.

\section*{Acknowledgment}
This work is supported by National Key Research and Development Program of China (Grant No. 2023YFB3904900).

\appendices

\section{Analysis of Lipschitz Continuous Condition}

\label{app:lipschitz analysis}

We start by considering the expression of the objective function $\boldsymbol{S}(\boldsymbol{X})$, which consists of a data fidelity term involving the observation operator $\mathcal{G}(\cdot)$, and several quadratic terms:
\[
\begin{split}
    \boldsymbol{S}(\boldsymbol{X}^{(k)}) =& \,\left\lVert \mathcal{G}(\boldsymbol{X}^{(k)})-\boldsymbol{Y_d}\right\rVert_2^2 \\ &+ \frac{\rho}{2}\left\lVert \boldsymbol{X}^{(k)} - \boldsymbol{Z}^{(k-1)}+\boldsymbol{V}^{(k-1)}\right\rVert_2^2.
\end{split}
\]

The quadratic terms possess constant second-order derivatives, and satisfy the Lipschitz continuity condition. 
The Lipschitz continuity of the Hessian of $\boldsymbol{S}(\boldsymbol{X})$ is primarily determined by the characteristics of the observation operator $\mathcal{G}(\cdot)$, which consists of three components: Fourier transforms, downsampling matrices, and Hadamard products. 
The contribution of each component is analyzed individually below.

\begin{itemize}
    \item \textbf{Fourier Transform}: This is a linear and unitary operation. Its derivative (Jacobian) is constant, and its second derivative (Hessian) is zero. Therefore, it satisfies the Lipschitz continuity condition trivially.

    \item \textbf{Downsampling Matrix}: The downsampling matrix used here performs random row selection from a full-size identity matrix, forming a binary selection operator. As a linear projection, it introduces no nonlinearity, thereby trivially satisfying the Lipschitz continuity condition.

    \item \textbf{Hadamard Product}: Since the element-wise (Hadamard) product involves a constant matrix determined by radar system parameters, it forms a diagonal linear operator with constant second-order derivatives, thus satisfying the Lipschitz continuity condition for the Hessian.
\end{itemize}

Therefore, there exists a Lipschitz constant \(L > 0\) such that for any \(\boldsymbol{X}\) and \(\boldsymbol{X}^{(k-1)}\),
\[
    \left\| \nabla^2 \boldsymbol{S}(\boldsymbol{X}) - \nabla^2 \boldsymbol{S}(\boldsymbol{X}^{(k-1)}) \right\| \leq L \left\| \boldsymbol{X} - \boldsymbol{X}^{(k-1)} \right\|.
\]

Since the Hessian of $\boldsymbol{S}(\boldsymbol{X})$ is Lipschitz continuous, its variation is bounded within a local neighborhood. When $\boldsymbol{X}$ is sufficiently close to $\boldsymbol{X^{(k-1)}}$, we have
\[
    \nabla ^2\boldsymbol{S}(\boldsymbol{X}^{(k-1)}) \left\lVert\boldsymbol{X} - \boldsymbol{X}^{(k-1)}\right\rVert \leq L\left\lVert\boldsymbol{X} - \boldsymbol{X}^{(k-1)}\right\rVert.
\]

In practical implementations, we adopt this estimated upper bound $L$ to approximate the second-order behavior of the objective function and derive Eq.~\ref{eq:admm taylor} accordingly.

\section{Proof of Eq.~\ref{eq:admm solution}}

\label{app:admm proof}

Eq.~\ref{eq:admm taylor} is to be simplified as follows.
\[
\begin{split}
    \boldsymbol{X}^{(k)} =& \,\underset{\boldsymbol{X}}{\arg\min}[\boldsymbol{S}(\boldsymbol{X}^{(k-1)})+\nabla\boldsymbol{S}(\boldsymbol{X}^{(k-1)})^H(\boldsymbol{X} - \boldsymbol{X}^{(k-1)}) \\
    &+\frac{L}{2}\left\lVert \boldsymbol{X} - \boldsymbol{X}^{(k-1)}\right\rVert_2^2]
\end{split}
\]

We combine the linear and quadratic terms by completing the square, as shown in the following equation.
\begin{equation}
    \begin{split}
    \boldsymbol{X}^{(k)} &= \,\underset{\boldsymbol{X}}{\arg\min}[\boldsymbol{S}(\boldsymbol{X}^{(k-1)})+\nabla\boldsymbol{S}(\boldsymbol{X}^{(k-1)})^H(\boldsymbol{X} - \boldsymbol{X}^{(k-1)}) \\
    &+\frac{L}{2}\left\lVert \boldsymbol{X} - \boldsymbol{X}^{(k-1)}\right\rVert_2^2 - \frac{1}{2L}\nabla\boldsymbol{S}(\boldsymbol{X}^{(k-1)})^H\nabla\boldsymbol{S}(\boldsymbol{X}^{(k-1)}) \\
    &+ \frac{1}{2L}\nabla\boldsymbol{S}(\boldsymbol{X}^{(k-1)})^H\nabla\boldsymbol{S}(\boldsymbol{X}^{(k-1)})]
    \end{split}
\end{equation}

After simplification, we obtain the following result: 
\begin{equation}
    \begin{split}
    \boldsymbol{X}^{(k)} =& \,\underset{\boldsymbol{X}}{\arg\min}[\frac{L}{2}\left\lVert \boldsymbol{X} - \boldsymbol{X}^{(k-1)} + \frac{1}{L}\nabla\boldsymbol{S}(\boldsymbol{X}^{(k-1)} \right\rVert_2^2 \\
    &+ \boldsymbol{S}(\boldsymbol{X}^{(k-1)}) - \frac{1}{2L}\nabla\boldsymbol{S}(\boldsymbol{X}^{(k-1)})^H\nabla\boldsymbol{S}(\boldsymbol{X}^{(k-1)})].
    \end{split}
\end{equation}
where $\boldsymbol{S}(\boldsymbol{X}^{(k-1)})$, $\frac{1}{2L}\nabla\boldsymbol{S}(\boldsymbol{X}^{(k-1)})^H\nabla\boldsymbol{S}(\boldsymbol{X}^{(k-1)})$ are constant terms irrelevant to the optimization variable $\boldsymbol{X}$.

The constant terms independent of $\boldsymbol{X}$ are omitted, as it does not affect the minimization, as
\begin{equation}
\boldsymbol{X}^{(k)} = \underset{\boldsymbol{X}}{\arg\min}\frac{L}{2}\left\lVert \boldsymbol{X} - \boldsymbol{X}^{(k-1)} + \frac{1}{L}\nabla\boldsymbol{S}(\boldsymbol{X}^{(k-1)}\right\rVert_2^2.
\end{equation}

By solving the minimization problem, we obtain an iterative solution in the form of a gradient descent step, as
\begin{equation}
\boldsymbol{X}^{(k)} =\boldsymbol{X}^{(k-1)} - \frac{1}{L}\nabla\boldsymbol{S}(\boldsymbol{X}^{(k-1)}).
\end{equation}

The gradient of the objective function $\boldsymbol{S}(\boldsymbol{X})$ is computed as follows. Since both $\mathcal{G}(\cdot)$ and $\mathcal{T}(\cdot)$ are composed of linear operations, they form a conjugate transpose pair~\cite{zhang2023nonsparse}, as
\[
\mathcal{G}^H(\cdot) = \mathcal{T}(\cdot).
\]
As a result, the first-order derivative (gradient) of $\mathcal{G}(\cdot)$ is given by $\mathcal{T}(\cdot)$, which simplifies the gradient expression.
\begin{equation}
\begin{split}
    \nabla\boldsymbol{S}(\boldsymbol{X}^{(k)}) =& \nabla\left\lVert \mathcal{G}(\boldsymbol{X}^{(k)})-\boldsymbol{Y_d}\right\rVert_2^2 \\
    &+ \frac{\rho}{2}\nabla\left\lVert \boldsymbol{X}^{(k)} - \boldsymbol{Z}^{(k-1)}+\boldsymbol{V}^{(k-1)}\right\rVert_2^2 \\
    =& \, 2\mathcal{G}^H[\mathcal{G}(\boldsymbol{X}^{(k)})-\boldsymbol{Y_d}] \\
    &+ \rho(\boldsymbol{X}^{(k)} - \boldsymbol{Z}^{(k-1)}+\boldsymbol{V}^{(k-1)}) \\
    =& \, 2\mathcal{T}[\mathcal{G}(\boldsymbol{X}^{(k)})-\boldsymbol{Y_d}] \\
    &+ \rho(\boldsymbol{X}^{(k)} - \boldsymbol{Z}^{(k-1)}+\boldsymbol{V}^{(k-1)}),
\end{split}
\end{equation}
where the imaging operator $\mathcal{T}(\cdot)$ is given by Eq.~\ref{eq:imaging operator}.

Substituting the gradient expression into the iterative solution, we have
\begin{equation}
\begin{split}
   \boldsymbol{X}^{(k)} =& \, (1-\frac{\rho}{L})\boldsymbol{X}^{(k-1)} + \frac{2}{L}\mathcal{T}[\boldsymbol{Y_d} - \mathcal{G}(\boldsymbol{X}^{(k-1)})] \\
   &+ \frac{\rho}{L}(\boldsymbol{Z}^{(k-1)} - \boldsymbol{V}^{(k-1)}).
\end{split}
\end{equation}

By redefining the parameters, the iterative solution can be simplified as follows, combining the previous estimate, the imaging residual, and the auxiliary variables:
\[
\begin{split}
   \boldsymbol{X}^{(k)} =& \, (1-\rho_n)\boldsymbol{X}^{(k-1)} + \mu\mathcal{T}[\boldsymbol{Y_d} - \mathcal{G}(\boldsymbol{X}^{(k-1)})] \\
   &+\rho_n(\boldsymbol{Z}^{(k-1)} - \boldsymbol{V}^{(k-1)}).
\end{split}
\]

\section{Complexity Analysis}

Each layer of ARSAR-Net consists of three modules: a reconstruction module, an adaptive regularizer module, and the scaled multiplier module.
The scaled multiplier module has linear computational complexity.
Therefore, we focus on analyzing the complexity of the other two modules. 

Assuming the scattering coefficient matrix $\boldsymbol{X}$ is of size $N \times N$. According to Eq.~\ref{eq:arsar-net module1}, this module involves linear operations, the imaging operator $\mathcal{T}$, and the observation operator $\mathcal{G}$, as defined in Eq.~\ref{eq:imaging operator} and Eq.~\ref{eq:inverse imaging operator}, respectively.
These operators consist primarily of 2D Fourier transforms and element-wise multiplications. Since the Fourier transform has complexity $\mathcal{O}(N^2 \log N)$ and the Hadamard product $\mathcal{O}(N^2)$, the overall complexity of the reconstruction module is
\[
    \mathcal{O}(N^2 logN) + \mathcal{O}(N^2) = \mathcal{O}(N^2 logN).
\]
Compared to conventional ADMM, which requires matrix inversion with $\mathcal{O}(N^3)$ complexity, the implementation without matrix inversion significantly reduces the computational cost.

As described in Sec.~\ref{sec:ada_reg}, ARSAR-Net consists of two variants, Swift and Pro, which differ in the architecture of their adaptive regularizer modules.

In the Swift variant, the complexity is dominated by convolutional layers, which require $\mathcal{O}(N^2 \times C_{in} \times C_{out} \times K^2)$. With the kernel size $K$ being negligible relative to the feature map size $N$, the convolution operation has a simplified complexity denoted as $\mathcal{O}(N^2 \times C_{in} \times C_{out})$. Despite the increased number of channels, the reduced spatial resolution maintains the overall complexity at $\mathcal{O}(N^2)$.
In contrast, this module in the Pro variant retains a full-resolution structure. While the size of feature maps remains fixed, the number of channels increases progressively. When the number of channels reaches its maximum, the input and output channels satisfy $C_{in} \times C_{out} \sim N$. Therefore, the complexity of this module is $\mathcal{O}(N^3)$.

Since the modules in ARSAR-Net are executed sequentially, the overall complexity is determined by the module with the highest complexity. Consequently, the computational complexity of ARSAR-Net Swift is
\[
    \mathcal{O}(N^2 logN) + \mathcal{O}(N^2) = \mathcal{O}(N^2 logN),
\]
while that of ARSAR-Net Pro is 
\[
    \mathcal{O}(N^2 logN) + \mathcal{O}(N^3) = \mathcal{O}(N^3).
\]

\bibliographystyle{IEEEtran}
\bibliography{main}

\end{document}